\documentclass[times,preprint]{aastex63}
\usepackage{amsmath}
\usepackage{comment}

\usepackage{color}
\usepackage{ulem}

\newtheorem{theorem}{Theorem}[section]

\received{}
\revised{}
\accepted{}

\submitjournal{ApJS}

\shorttitle{High Dimensional Statistical Analysis of NGC~253}
\shortauthors{Takeuchi et al.}
\graphicspath{{./}{figures/}}
\begin{document}

\title{High Dimensional Statistical Analysis and its Application to ALMA Map of NGC 253}

\correspondingauthor{Tsutomu T. {Takeuchi}}
\email{tsutomu.takeuchi.ttt@gmail.com}

\author[0000-0001-8416-7673]{Tsutomu T.\ Takeuchi}
\affiliation{Division of Particle and Astrophysical Science, Nagoya University, Furo-cho, Chikusa-ku, Nagoya 464--8602, Japan}
\affiliation{The Research Center for Statistical Machine Learning, the Institute of Statistical Mathematics, 10-3 Midori-cho, Tachikawa, Tokyo 190---8562, Japan}
 
\author[0000-0001-6814-4463]{Kazuyoshi Yata}
\affiliation{Institute of Mathematics, University of Tsukuba, 1--1--1 Tennodai, Tsukuba, Ibaraki 305--8571, Japan}

\author{Kento Egashira}
\affiliation{Department of Information Sciences, Tokyo University of Science, 2641 Yamazaki, Noda, Chiba 278--8510, Japan}
\affiliation{Graduate School of Science and Technology, University of Tsukuba, 1--1--1 Tennodai, Tsukuba, Ibaraki 305--8571, Japan}

\author{Makoto Aoshima}
\affiliation{Institute of Mathematics, University of Tsukuba, 1--1--1 Tennodai, Tsukuba, Ibaraki 305--8571, Japan}

\author{Aki Ishii}
\affiliation{Department of Information Sciences, Tokyo University of Science, 2641 Yamazaki, Noda, Chiba 278--8510, Japan}

\author[0000-0002-9217-1696]{Suchetha Cooray}
%\affiliation{Division of Particle and Astrophysical Science, Nagoya University, Furo-cho, Chikusa-ku, Nagoya 464--8602, Japan}
\affiliation{National Astronomical Observatory of Japan, National Institutes of Natural Sciences (NINS), 2-21-1 Osawa, Mitaka, Tokyo 181-8588, Japan}
\affiliation{Research Fellow of the JSPS (PD)}

\author{Kouichiro Nakanishi}
\affiliation{National Astronomical Observatory of Japan, National Institutes of Natural Sciences (NINS), 2-21-1 Osawa, Mitaka, Tokyo 181-8588, Japan}
\affiliation{Department of Astronomy, School of Science, Graduate University for Advanced Studies (SOKENDAI), 2-21-1 Osawa, Mitaka, Tokyo 181-8588, Japan}

\author[0000-0002-4052-2394]{Kotaro Kohno}
\affiliation{Institute of Astronomy, Graduate School of Science, The University of Tokyo
2-21-1, Osawa, Mitaka, Tokyo, 181-0015, Japan}

\author{Kai T.\ Kono}
\affiliation{Division of Particle and Astrophysical Science, Nagoya University, Furo-cho, Chikusa-ku, Nagoya 464--8602, Japan}

\begin{abstract}
In astronomy, if we denote the dimension of data as $d$ and the number of samples as $n$,  we often meet a case with $n \ll d$. 
Traditionally, such a situation is regarded as ill-posed, and there was no choice but to throw away most of the information in data dimension to let $d < n$. 
The data with $n \ll d$ is referred to as high-dimensional low sample size (HDLSS). 
{}To deal with HDLSS problems, a method called high-dimensional statistics has been developed rapidly in the last decade.
In this work, we first introduce the high-dimensional statistical analysis to the astronomical community.
We apply two representative methods in the high-dimensional statistical analysis methods, the noise-reduction principal component analysis (NRPCA) and 
automatic
sparse principal component analysis (A-SPCA), to a spectroscopic map of a nearby archetype starburst galaxy NGC~253 taken by the Atacama Large Millimeter/Submillimeter Array (ALMA). 
The ALMA map is a typical HDLSS dataset. 
First we analyzed the original data including the Doppler shift due to the systemic rotation.
The high-dimensional PCA could describe the spatial structure of the rotation precisely.
We then applied to the Doppler-shift corrected data to analyze more subtle spectral features.
The NRPCA and R-SPCA could quantify the very complicated characteristics of the ALMA spectra.
Particularly, we could extract the information of the global outflow from the center of NGC~253.
This method can also be applied not only to spectroscopic survey data, but also any type of data with small sample size and large dimension. 
\end{abstract}

\keywords{methods: statistical --- ISM: lines and bands --- galaxies: ISM ---galaxies: star formation --- galaxies: individual (NGC~253) --- radio lines: galaxies}
%\keywords{Astrostatistics techniques (1886); Starburst galaxies (1570); Molecular spectroscopy (2095); Molecular gas (1073); Millimeter astronomy (1061); Submillimeter astronomy (1647); Radio interferometry (1346)}

\section{Introduction} \label{sec:intro}

In the past decade, the term ``Big Data'' has been widely recognized and observed almost all the field of scientific researches, of course also in astronomy. 
Current astronomical instruments provide us with overwhelmingly large data, providing quantitative information of their physical condition. 
The most typical big data in astronomy is the product from large surveys, such as 
the Sloan Digital Sky Survey (SDSS)\footnote{\tt https://www.sdss.org/surveys/.}, 
the Dark Energy Spectroscopic Instrument (DESI) Survey\footnote{\tt https://www.desi.lbl.gov/.}, 
Gaia\footnote{\tt https://www.cosmos.esa.int/web/gaia.}, 
the Panoramic Survey Telescope And Rapid Response System (Pan-STARRS) Surveys\footnote{\tt %https://panstarrs.ifa.hawaii.edu/pswww/.}, 
https://outerspace.stsci.edu/display/PANSTARRS/.},
among many others. 

We should note, however, that there is another type of big data in astronomy. 
Detailed observations are often very time-consuming, and it is not easy to map objects to obtain many independently sampled measurements. 
Such a situation is frequently found in the integral field spectroscopy (e.g., 
UVES at the Very Large Telescope (VLT)\footnote{\tt https://www.eso.org/sci/facilities/paranal/instruments/uves.html.},
SAURON on the William Herschel Telescope\footnote{\tt %https://www.ing.iac.es/PR/wht$\_$info/whtsauron.html.},
https://www.ing.iac.es/PR/SH/SH2007/sauron.html.}, 
SDSS MaNGA data\footnote{\tt https://www.sdss.org/surveys/manga/.}, among others. 
Another typical example of this kind is spectroscopic mapping surveys by radio interferometers (e.g., Atacama Large Millimeter/Submillimeter Array) or by satellite instruments at X-ray, ultraviolet or mid/far-infrared wavelengths. 
In such a case, if we denote the dimension in the wavelength (or frequency) with $d$ and the number of samples with $n$,  we often find that $n \ll d$. 

Classical statistical analysis implicitly assumed a sample size $n$ is much larger than data dimension $d$ ($n \gg d$). 
As we mentioned, this is not true for many cases in scientific researches.
Traditionally in astrophysics, such a situation is regarded as an ill-posed problem, and there was no choice but to throw away most of the information in wavelength direction to let $d < n$. 
It was more or less the same for other research fields before 2010.
Many researchers have simply given up further analysis, or even did not find a need to deal with such problems before the advent of the Big Data era. 
For example, \citet{Hand1994} published a well-known compilation of statistical data, containing more than 500, but most of them satisfied the classical condition of the assumption $d < n$. 
However, in late 1990's, data with $d \gg n$ have suddenly started to appear along with the development of the information science.
One of the epoch-making example of such kind of researches was \citet{Golub531}.
They have used the gene expression monitoring by DNA microarrays applied to human acute leukemias as a test case. 
The dimension of the DNA microarray was $d=7129$ and the sample size $n = 72$. 
Statistics in 1990's could not guarantee the accuracy for the estimation in such a case, since it was on the basis of $d < n$. 

Various fundamental problems were known for data with the condition of $n \ll d$: 
\begin{enumerate}
    \item Sample covariance matrix is usually numerically unstable or does not exist, and classical multivariate analysis cannot work. 
    \item We meet various type of ``curse of dimensionality''. 
    \item Calculation cost is often unreachably tremendous. 
\end{enumerate}
A method to handle such a problem have been desired since then. 
In order to study such type of data with a very high dimension and small sample data, first of all we had to push back the boundaries of the traditional technique of multivariate statistical analysis. 
This is because traditional methods are not able to capture the characteristics of the high-dimensional data space and hence, not able to make use of rich information which the data intrinsically contain.   

The situation has drastically changed in 2000's. 
A turning point was brought from the field of probability theory and theoretical physics based on the random matrix theory. 
In these studies, the asymptotic behavior of eigenvalues of the sample covariance matrix in the limit as $d \longrightarrow \infty$ was explored under Gaussian or Gaussian-type assumptions on the population distribution when $d$ and $n$ increase at the same rate, i.e., $n/d \longrightarrow c > 0$ ($c$: const.) \citep[e.g.,][]{baik2005,BAIK20061382,johnstone2001,10.2307/24307692}. 
In our current interest, however, the assumption as $n/d \longrightarrow c > 0$ and Gaussian(-type) are actually unrealistically restrictive. 
A seminal paper by \citet{Hall2005} introduced a new concept, high dimension low sample size (HDLSS) asymptotics\footnote{The first consideration on the HDLSS asymptotics, however, dates back to 1980's by the work of \citet{Casella1982}}. 
The limits they have studied were taken as the dimension of the data $d \longrightarrow \infty$, while the sample size $n$ was fixed. 
At that time, this was a revolutionary departure from the traditional asymptotics based on $n \longrightarrow \infty$ with $d$ fixed. 

\citet{Hall2005} and \citet{10.2307/20441411} discussed an important representation of the HDLSS data, but still under the assumption of Gaussian(-type). 
In contrast, \citet{YATA2012193} developed a new theory by clearing away this assumption, and discovered a completely different type of representation when the Gaussian assumption is not satisfied (see \S~\ref{sec:high_dimension} for the details). 
These works were the foundation of research in mathematical statistics on very high dimensional data analysis, and evolved into an important part of the currently fashionable research area, the Big Data Science. 
The new insight shed light to various areas of studies as genomics, medical research, neuroscience, and image and shape analysis. 
Then, in their sequence of papers, M.\ Aoshima, K.\ Yata, A.\ Ishii and collaborators further developed a new framework of statistical methodology to tackle this type of problems \citep[e.g., ][and references therein]{doi:10.1080/07474946.2011.619088,YATA2012193, YATA2013334,Aoshima2014,doi:10.1080/07474946.2015.1063256,ISHII2016186,aoshima2019}. 
Readers who are interested in the mathematical background of the high-dimensional statistical analysis itself are guided to some reviews,\citep[e.g.,][]{aoshima2017,Aoshima2018}. 

In spite of the great developments in other research fields, this new methodology of statistics seemed to be completely unknown in astronomy.
High dimensional data as the ALMA data are regarded as a sum of informative signals and noise, but we cannot separate them quantitatively without prior information. 
Supposing that we know the frequencies of emission lines is a kind of such prior information which has been traditionally used in traditional astronomy, but this is not what we prefer. 
While some frequency regions are considered to contain no information, we proceeded extracting information not by traditional procedure, but by an objective statistical method without a prior information. 
The high-dimensional statistical analysis is one of such an objective methods.

Thus, in this work, we introduce the method of high-dimensional statistical analysis for the HDLSS data in astronomy. 
In order to examine the performance of the high-dimensional methods to the astronomical data, we apply it to a spectroscopic map of the central region of NGC~253.
NGC~253 is a nearby prototypical starburst galaxy \citep{1980ApJ...238...24R}, and has been observed in various wavelengths. 
We apply the high-dimensional principal component analysis (PCA) to the ALMA spectral map \citep[][hereafter A17]{2017ApJ...849...81A}. 
A17\nocite{2017ApJ...849...81A} have presented an $8~\mbox{pc} \times 5~\mbox{pc}$ resolution view of the central $\sim 200$~pc region of NGC~253, based on the ALMA Band~7 observations. 
As we introduce in details, the ALMA maps can be regarded as being typically HDLSS. 
The spectrum of NGC~253 is very rich in molecular lines \citep[e.g.,][]{2017ApJ...849...81A,2021A&A...656A..46M}. 
Another focus of this work is to classify the spectra full of spectral line features by the PCA. 
Evolution of galaxies is mainly driven by the star formation, a transition from interstellar medium (ISM) to stars. 
Various phases of the ISM are related, and the evolution of the ISM is a key to complete the understanding of the galaxy evolution. 
Spectroscopic observations are of vital importance to extract and interpret the information of matter in galaxies\citep[e.g.,][]{2021A&A...656A..46M}. 
Then, this classification is potentially a powerful tool for the studies of galaxy evolution if it turns out to be useful. 

This paper is organized as follows. 
In \S~\ref{sec:high_dimension}, we try to introduce the concept of HDLSS data and high-dimensional statistical analysis, with an emphasis of the high-dimensional principal component analysis (PCA). 
There we present some important theorems that characterize the surprising features of the HDLSS data, without detailed mathematical proofs. 
Then in \S~\ref{sec:data}, we move on to the actual analysis of astronomical data. 
As already mentioned, we used an ALMA map of NGC~253. 
We summarize the data properties in this section. 
Our data analysis is twofold: first we examine if the high-dimensional statistical analysis is really useful for the analysis of spectroscopic mapping data in \S~\ref{sec:preparatory_analysis}. 
After confirming its excellent performance in this type of analysis, we further go to a deeper analysis of the NGC~253 molecular lines in \S~\ref{sec:results_discussion}.
\S~\ref{sec:conclusion} is devoted to the conclusions of this work.
Some technical details related to both theoretical and practical methods in this work are presented in Appendix. 
Appendices~\ref{sec:dual_covariance} and \ref{sec:geometric_representation} describe the properties of the dual sample covariance and its geometric representation.
In Appendix~\ref{sec:non_gaussian}, we present the methodology for non-Gaussian data that was not discussed in the main text.
Comparison between the results of the traditional and high-dimensional PCAs is shown in Appendix~\ref{sec:comparison_PCA}. 
The method to correct the Doppler shift is presented in Appendix~\ref{sec:Doppler_shift_correction}.

\section{Method: High-Dimensional Statistical Analysis}\label{sec:high_dimension}

\subsection{High-Dimensional Statistical Analysis}

The HDLSS data have a sample size $n$ and data dimension $d$ with $n \ll d$. 
In this work, we mean the newly developed statistical methodology to deal with the HDLSS data by ``high-dimensional statistical analysis''.  

\subsubsection{Unusual behavior of the high-dimensional statistics}
\label{unusual}

As a pedagogical example, we demonstrate a typical ``unusual'' properties only appear in the high-dimensional statistical analysis, known as the strong inconsistency. 
We consider a sample mean $\overline{\vec{x}}$, 
\begin{eqnarray}\label{eq:sample_mean}
  \overline{\vec{x}}_n \equiv \frac{1}{n} \sum _{i=1}^{n} \vec{x}_i \; .
\end{eqnarray}
Suppose that $\vec{x}_i$s are drawn from a $d$-dimensional distribution with a mean $\vec{\mu}$. 
Under the usual condition often supposed in the traditional statistics, $d/n \longrightarrow 0$, we have
\begin{eqnarray}\label{eq:consistency}
  \left\|  \overline{\vec{x}} - \vec{\mu} \right\| \overset{\mbox{P}}{\longrightarrow} 0 \; 
\end{eqnarray}
where the superscript P stands for the convergence in probability\footnote{More precisely, $\underset{d/n \rightarrow 0 }{\lim }\mathbb{P}\left(\left\|\vec{x}_n - \vec{\mu} \right\| > \epsilon \right) = 0$ for any $\epsilon > 0$. 
Here $\| \cdot \|$ stands for the Euclid norm, and $\mathbb{P}(A)$ is the probability of an event $A$. 
Similarly, $\underset{d/n \rightarrow \infty }{\lim }\mathbb{P}\left(\left\|\vec{x}_n - \vec{\mu} \right\| < \epsilon \right) = 0$ for any $\epsilon > 0$ for eq.~(\ref{eq:inconsistency}). 
}. 
Equation~(\ref{eq:consistency}) is referred to as the consistency of the sample average. 
However, in the high-dimensional statistics, $d/n \longrightarrow \infty$, and shockingly we have
\begin{eqnarray}\label{eq:inconsistency}
  \left\| \overline{\vec{x}} - \vec{\mu} \right\| \overset{\mbox{P}}{\longrightarrow} \infty \; .
\end{eqnarray}
This is the so-called {\sl strong inconsistency} in the HDLSS framework. 
What eq.~(\ref{eq:inconsistency}) mean is that the noise in the observed data drastically increases with the increase of the dimension $d$. 
Traditional statistical techniques suppose that the noise is negligible compared to the data information. 
Hence, the traditional statistics can never handle the HDLSS data. 

In order to have more concrete image on such ``counter-intuitive'' phenomena, we examine the behavior of this huge noise in high dimension. 
For this, however, some mathematical background is necessary. 
In the following we introduce the background mathematical concepts. 
We do not try to introduce them in a strict manner, either not try to give rigorous mathematical proofs, even if the following sections seem to be rather mathematically inclined.

\subsubsection{{Dual sample covariance}}

%\subsubsection{Geometric representation of the data}

Consider a parent distribution of dimension $d$, with a mean vector $\vec{\mu} = \vec{0}$ and a covariance matrix $\tilde{\Sigma}_d > \tilde{O}$ (i.e., all the eigenvalues are positive).
We can adopt this constraint for the sake of simplicity. 
%without a loss of generality. 
Let $\lambda_i \; (i = 1, \dots, d)$ be the eigenvalues of $\tilde{\Sigma}$, and sort them out as 
\begin{eqnarray}
  \lambda_1 \geq \lambda_2 \geq \dots \geq \lambda_d\ (>0) \; .      
\end{eqnarray}
Then, we can make an eigenvalue decomposition of
\begin{eqnarray}
  \tilde{\Sigma}_d &=& \tilde{H} \tilde{\Lambda} \tilde{H}^\top \\ 
  \tilde{\Lambda} &\equiv&  \mbox{diag}\left( \lambda_1, \lambda_2, \dots  \lambda_d \right) \; .
\end{eqnarray}
Here, $\tilde{H} = \left[\vec{h}_1, \dots, \vec{h}_d \right]$ is an orthogonal matrix of the corresponding  eigenvectors. 
Note that these are all outrageously huge matrices. 

\begin{figure*}[tb]
    \centering
    \includegraphics[width=0.48\textwidth]{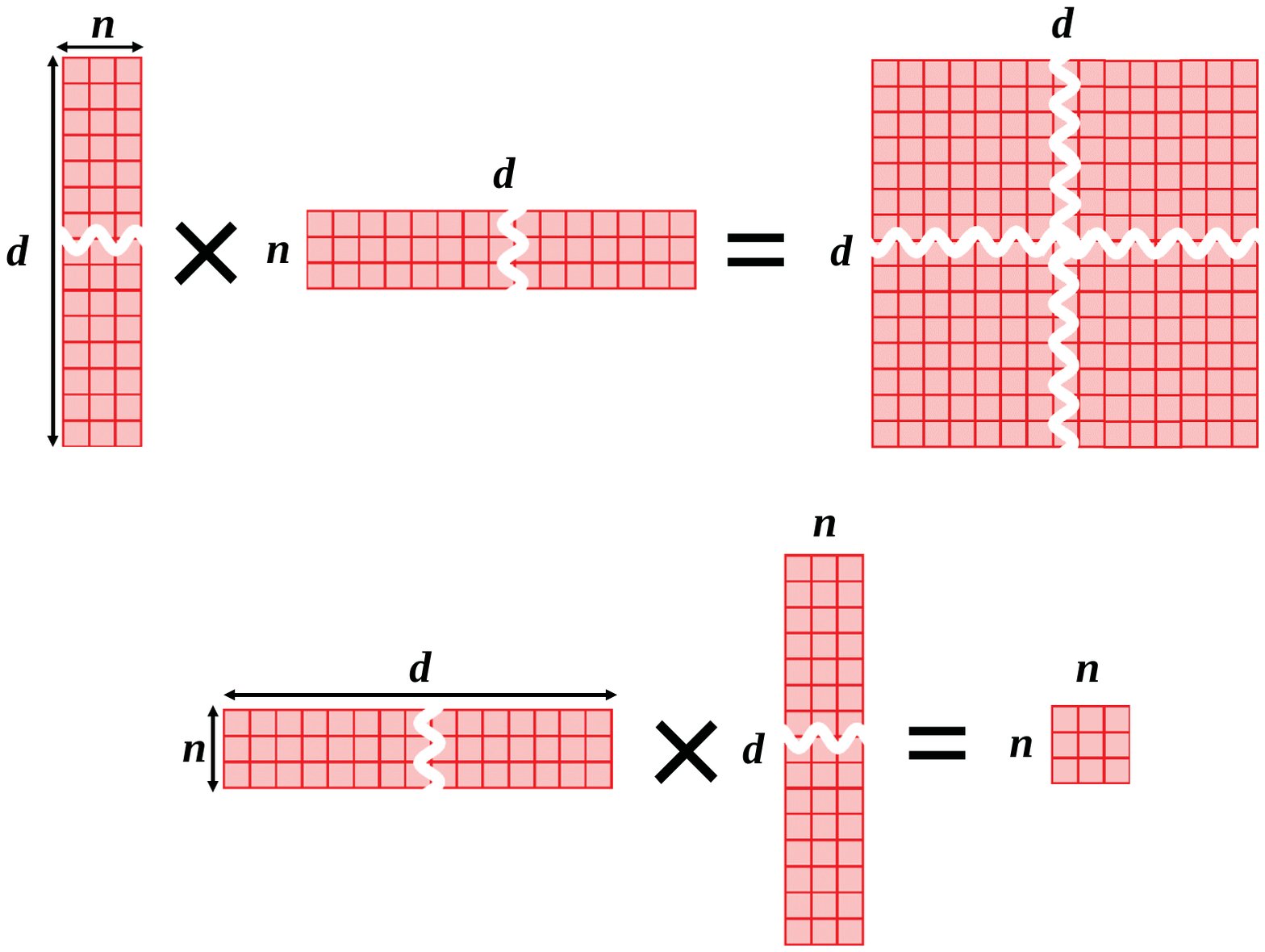}
    \caption{Schematic description of the relation between a sample covariance matrix $\tilde{S}$ and its dual matrix $\tilde{S}_{\rm D}$. 
    Top panel describes the sample covariance matrix, and Bottom panel is the dual matrix, respectively. 
    }\label{fig:sample_dual_matrices}
\end{figure*}

Now we draw a set of $n$ samples from this parent population ($d > n$), $\vec{x}_1, \vec{x}_2, \dots, \vec{x}_n$.
Suppose that we describe this set of data by a $(d \times n)$ data matrix
\begin{eqnarray}
  \tilde{X} = \left( \vec{x}_1, \vec{x}_2, \dots, \vec{x}_n \right)\;, 
\end{eqnarray}
and $\vec{x}_i \; (i=1, \dots, n)$ are independently and identically distributed (i.i.d.). 
{We should note that the i.i.d.\ condition is not always satisfied in practice. 
However, this condition is often required for the rigorous construction of the mathematical methodology}.  
The sample covariance matrix ($d \times d$) is expressed as
\begin{eqnarray}
  \tilde{S} = \frac{1}{n} \tilde{X}\tilde{X}^\top \;.
\end{eqnarray}
Here, we define the dual sample covariance ($n \times n$), 
\begin{eqnarray}
  \tilde{S}_{\rm D} \equiv \frac{1}{n} \tilde{X}^\top \tilde{X} \; .
\end{eqnarray}
A schematic description of a sample covariance matrix $\tilde{S}$ and its dual matrix $\tilde{S}_{\rm D}$ is shown in Fig.~\ref{fig:sample_dual_matrices}. 
If we consider the sample eigenvalues of $\tilde{S}_{\rm D}$, its $n$ eigenvalues 
\begin{eqnarray}
  \hat{\lambda}_1 \geq \hat{\lambda}_2 \geq \dots \geq \hat{\lambda}_n > 0       
\end{eqnarray}
are common with the first $n$ eigenvalues of $\tilde{S}$. 
See Appendix \ref{sec:dual_covariance} for the merit of 
$\tilde{S}_{\rm D}$. 

An important piece of the initial exploration of HDLSS asymptotics in \citet{Hall2005} was the revelation of a surprisingly rigid geometric structure that naturally emerges from the asymptotics in an increasingly random setting, referred to as the geometric representation by them. 
Recall the eigenvalue decomposition of the covariance matrix $\tilde{\Sigma} = \tilde{H} \tilde{\Lambda} \tilde{H}^\top$.  
%If we denote $\tilde{H} = \left[\vec{h}_1, \dots, \vec{h}_d \right]$ is an orthogonal matrix of the corresponding eigenvectors. 
Let 
\begin{eqnarray}
  \tilde{X} \equiv \tilde{H} \tilde{\Lambda}^\frac{1}{2} \tilde{Z}.
\end{eqnarray}
Then, $\tilde{Z}$ is a $(d \times n)$ sphered data matrix from a distribution with the identity covariance matrix. 
Here, we write $\tilde{Z} = \left[\vec{z}_1, \dots, \vec{z}_d \right]^\top$ and $\vec{z}_i = (z_{i1}, \dots,  z_{in})^\top, \;  (i = 1, \dots, d)$. 
Note that $\mathbb{E}(z_{ij} z_{i'j}) = 0 \; (i \neq i')$ and $\mathbb{V}(\vec{z}_i) = \tilde{I}_n$, where $\mathbb{E}$ and $\mathbb{V}$ denote the expectation and variance of a random variable, and $\tilde{I}_n$ is the $n$-dimensional identity matrix. 
We assume that the fourth moments of each variable in $\tilde{Z}$ are uniformly bounded. 
Note that if $\tilde{X}$ is Gaussian, the $z_{ij}$s are i.i.d. standard normal random variables.
We also define the eigenvalue decomposition of $\tilde{S}_{\rm D}$ by 
\begin{eqnarray}
 \tilde{S}_{\rm D} = \sum _{i = 1}^n  \hat{\lambda}_i \vec{u}_i \vec{u}_i^\top \; .
\end{eqnarray}
Here, $\vec{u}_i$ denotes
a unit eigenvector corresponding to $\hat{\lambda}_i$. 
{
See Appendix \ref{sec:geometric_representation} for details of geometric representations of $\tilde{S}_{\rm D}$. 
}
Let $\hat{\vec{h}}_i$ be the $i$-th eigenvector of $\tilde{S}$. 
Note that $\hat{\vec{h}}_i$ can be calculated by
\begin{eqnarray}\label{eq:vector}
\hat{\vec{h}}_i=(n\hat{\lambda} _i)^{-1/2} \tilde{X} {\vec{u}}_i. 
\end{eqnarray}

\subsection{High-Dimensional Principal Component Analysis}

We adopt the same setting as the previous sections for the data distribution: we suppose a random variable with a mean $\vec{\mu} = \vec{0} $ and a covariance matrix  $\tilde{\Sigma}_d > 0$. 
%, and the fourth order moment of $Z$ is uniformly bounded. 
We assume the following model, referred to as ``the generalized spiked model'', for the eigenvalue distribution of $\tilde{\Sigma}_d$,
\begin{eqnarray}\label{eq:spike_model}
  \lambda_i = \left\{
  \begin{array}{ll}a_i d^{\alpha_i} & \quad\mbox{for }\quad i = 1,\dots ,  m \\
  c_i & \quad\mbox{for } \quad i = m + 1, \dots , d \; 
  \end{array}
  \right.
\end{eqnarray}
where $a_i(> 0), c_i(> 0)$ and $\alpha_i \; (\alpha_ 1 \geq \dots \geq \alpha_ m > 0)$ are unknown constants that preserves the order $\lambda_ 1 \geq \dots \geq \lambda_ d$, and $m$ is an unknown positive fixed integer\footnote{One may feel suspicious if such a restrictive model can really describe the general properties of eigenvalue distribution of real data. 
However, as we see in \S\S~\ref{sec:preparatory_analysis} and \ref{sec:results_discussion}, the generalized spiked model really well represents the actual eigenvalues.}. 
{The $m$ stands for the number of spiked eigenvalues in the sense of quadratic contribution ratio. }
The details of the spiked model including the generalized spiked model was considered and given in \cite{YATA2013334}.
The eigenvalues $\lambda_i \; (i \leq m)$ describe the intrinsic information of the data, and $\lambda_i \; (i \geq m+1)$ represent the noise.
Note that 
\begin{eqnarray}
  \frac{\displaystyle \sum _{i = m+1}^{d} \lambda_i^2}{\displaystyle \left( \sum _{i = m+1}^{d} \lambda_i \right)^2 } \overset{\rm P}{\longrightarrow} 0 \quad \mbox{as} \quad d \longrightarrow \infty 
\end{eqnarray}
i.e., the noise satisfies the sphericity condition [eq.~(\ref{eq:sphericity_condition})].
We set the sample covariance matrix $\tilde{S} = n^{-1} \tilde{X}\tilde{X}^\top$ as above. 
Here, we define $n(d)$ as the size of a sample depending on $d$. 

\subsubsection{Limitation of the traditional principal component analysis (PCA)}

\citet{doi:10.1080/03610910902936083} investigated the properties of the traditional PCA when it is applied to HDLSS data, and demonstrated the condition so that the estimation by the traditional PCA would have the consistency in the form of $n(d) = d^\gamma$ \footnote{In this context, consistency stand for the situation as follows: 
Consider an estimator $t(n)$ ($n$: sample size) is a consistent estimator for a parameter $\theta$ if and only if, for all $\epsilon > 0$, we have
\begin{eqnarray*}
  \lim _{n\to \infty } \mathbb{P} \left\{\left|t(n)-\theta \right|<\epsilon \right\}=1 \, .
\end{eqnarray*}
} for a positive constant $\gamma$ and the limitation of the traditional PCA. 
They have given the following two theorems on the eigenvalues, depending if the condition would satisfy 
\begin{eqnarray}\label{eq:independence}
  z_{ij} \quad (i = 1, \dots, d, \; j = 1, \dots, n) \mbox{ are independent. }
\end{eqnarray}

{\theorem{\bf [\citet{doi:10.1080/03610910902936083} I]}\label{th:YAI}\\
If $Z$ satisfies Condition (\ref{eq:independence}), for $i \leq m$, 
\begin{eqnarray}\label{eq:eigenvalues}
  \frac{\hat{\lambda}_i}{\lambda_i} = 1 + o_{\rm P} (1) \; , 
\end{eqnarray}
under the conditions 
\begin{enumerate}
    \item $d \longrightarrow \infty$ and $n \longrightarrow {\infty}$ for $i$ that satisfies $\alpha_i > 1$ .  
    \item $d \longrightarrow \infty$ and $d^{1 - \alpha_i}/n(d) \longrightarrow 0$ for $i$ that satisfies $\alpha_i \in (0, 1]$.
\end{enumerate}
}
\noindent
{
See Appendix \ref{sec:non_gaussian} for details of the consistency 
when $Z$ does not satisfy Condition (\ref{eq:independence}). 
}

{Theorem~\ref{th:YAI} gives the condition that the estimation by the traditional PCA would have the consistency. 
The condition described by $d \longrightarrow \infty$ and $n \longrightarrow \infty$ in (1) of these theorems is a mild condition in the sense that one can choose $n$ free from $d$ 
or that $n$ may be much smaller than $d$ such as $n=\log{d}$. 
However, in (2), $n$ is heavily depending on $d$.} 
Hence, we should be very careful to apply the conventional PCA to the HDLSS data.
In order to overcome this fundamental problem of the conventional PCA, \citet{120002381162} and YA12\nocite{YATA2012193} proposed two new PCAs for HDLSS data. 
In this work, we focus on one of them, the noise-reduction methodology (YA12\nocite{YATA2012193}). 

\subsubsection{Noise-reduction PCA (NRPCA): a new method for HDLSS data}

\begin{figure*}[tb]
    \centering
    \includegraphics[width=0.9\textwidth]{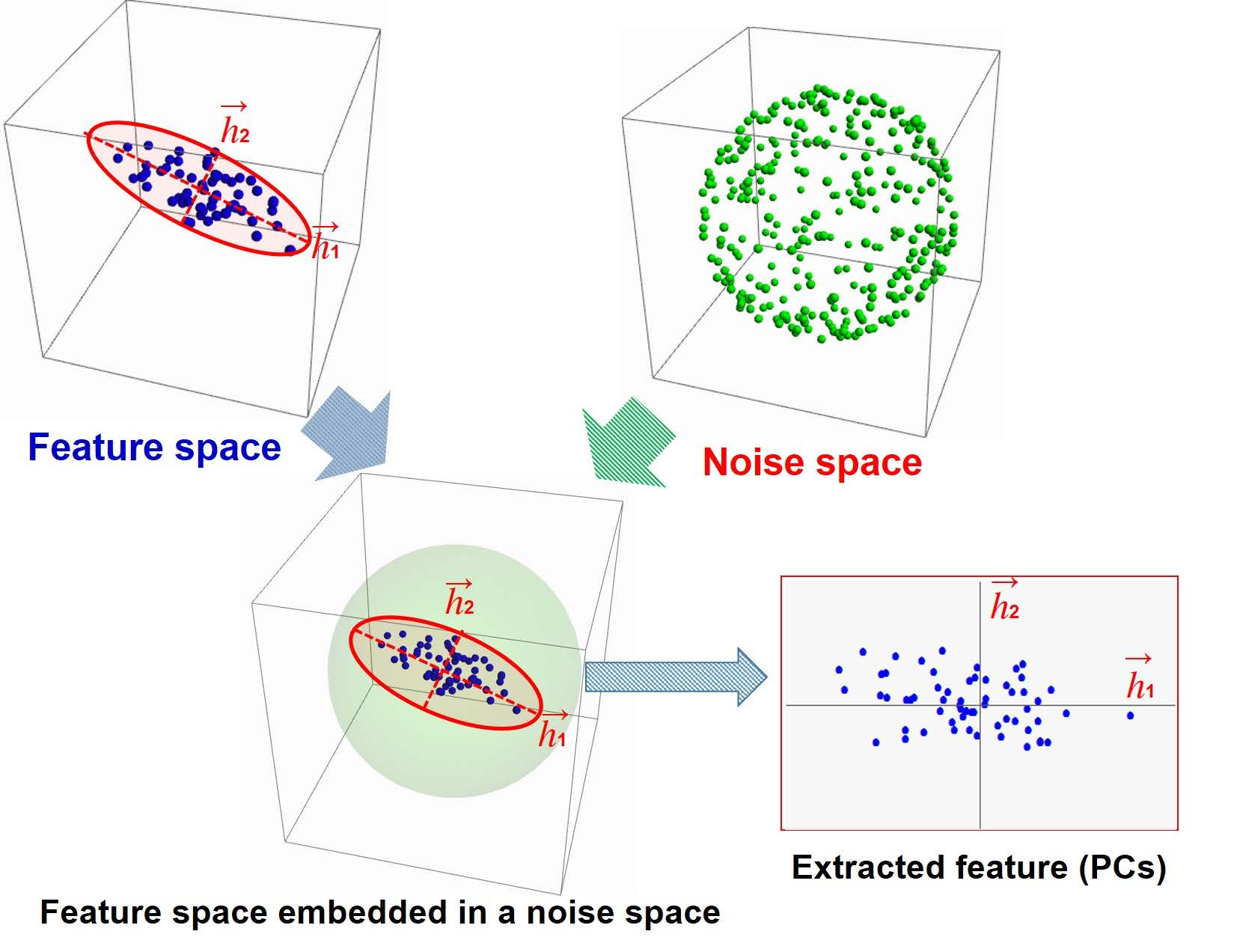}
    \caption{A schematic description of the high-dimensional principal component analysis (PCA). 
    Even if we have significant important features, because of the existence of a huge noise sphere, they are completely embedded in the observed data. 
    The high-dimensional PCA can subtract the noise sphere and enables to extract the desired features. 
    }\label{fig:high_dimension_PCA}
\end{figure*}

Here we introduce the high-dimensional PCA.
First we informally present the concept of the high-dimensional PCA, which is schematically described in Fig.~\ref{fig:high_dimension_PCA}. 
Because of the enormous noise that concentrates on the sphere in the data space, the intrinsically important characteristics of the data are completely buried under the noise sphere and obscured. 
However, it is not hopeless to estimate the characteristics even under such a situation, thanks to the fact that the noise behavior around the sphere can be well analyzed and essentially {\sl subtracted}. 
The method to concretely realize this idea is the noise-reduction (NR) methodology proposed by YA12\nocite{YATA2012193}. 

We try to explain it in a more rigorous manner. 
YA12\nocite{YATA2012193} proposed the NR methodology derived from the geometric representation [eq.~(\ref{eq:ahn_jung})]. 
%eq:sphericity_condition
%})]. 
First we decompose $\tilde{S}_{\rm D}$ as 
\begin{eqnarray}
  n\tilde{S}_{\rm D} = \sum  _{i=1}^m \lambda_i \vec{z}_i \vec{z}_i^\top + \sum _{i=m+1}^d\lambda_i  \vec{z}_i \vec{z}_i^\top 
\end{eqnarray}
{
The second term 
represents the noise.
%The second term representing the noise is expressed as
%\begin{eqnarray}
%  \frac{\displaystyle \sum_{i=m+1}^d \lambda_i^2  }{\displaystyle \left( \sum_{i = m+1}^d  \lambda_j\right)^2} \longrightarrow 0 \quad 
%  \mbox{as} \quad  d \longrightarrow \infty  \; .
%\end{eqnarray}
Then, from Theorem~\ref{th:YA2012}, 
under Condition~(\ref{eq:independence})}, we obtain the following geometric representation for the noise
\begin{eqnarray}
  \frac{\displaystyle\sum _{i = m+1}^d  \lambda_j\vec{z}_i \vec{z}_i^\top}{\displaystyle \sum _{i = m+1}^d  \lambda_j} \overset{\rm P}{\longrightarrow} \tilde{I}_n \quad \mbox{as} \quad d \longrightarrow \infty ,
\end{eqnarray}
i.e., the noise part becomes deterministic under the HDLSS condition.
This result enables us to remove the noise part and obtain the formula
\begin{eqnarray}
  \check{\lambda}_i = \hat{\lambda}_i - \frac{1}{\displaystyle n - i} \left(\mbox{tr}\,\tilde{S}_{\rm D} - \sum_{j = 1}^i \hat{\lambda}_i \right) \qquad (i = 1, \dots, n-1) \; . \label{eq:eigenvalues_estimation}
\end{eqnarray}
Estimation of $m$ is given by \citet{AY18}. 
We have the following theorem.
{\theorem{\bf [\citet{YATA2012193} II]}\label{th:YA2012b}. \\
When the components of $\tilde{Z}$ satisfy Condition~(\ref{eq:independence}), it holds that for
$i \leq   m$, 
\begin{eqnarray}
\frac{\check{\lambda}_i}{\lambda_i} = 1 + o_{\rm P}  (1) \; , 
\end{eqnarray}
under the conditions that
\begin{enumerate}
    \item $d \longrightarrow \infty$ and $n \longrightarrow \infty$ for $i$ such that $\alpha_i  > 1/2$, 
    \item $d \longrightarrow \infty$ and $d^{1 - 2\alpha_i}/n(d) \longrightarrow 0$ for $i$ such that $\alpha_i \in (0, 1/2]$. 
\end{enumerate}
}

\noindent
If we compare Theorem~\ref{th:YA2012b} with Theorem~\ref{th:YAI}, we see that the consistency holds for $\check{\lambda}_i$s under milder conditions than $\hat{\lambda}$ [under Condition~(\ref{eq:independence})] 
\footnote{{If $\alpha_i  > 1/2$, the NR methodology has the consistency even when $n$ is much smaller than $d$ such as $n=\log{d}$.} }
. 
%%%
\begin{comment}
%
{\theorem{\bf [\citet{YATA2012193} III]}\label{th:YA2012c}\\ 
Let $\mathbb{V} (z_{ij}^2) \equiv M_i (> 0)$ for $i = 1, \dots , d \; (j = 1, \dots , n)$.
Assume that $\lambda_i  \; (i \leq m)$ has multiplicity one. When the components of $\tilde{Z}$ satisfy Condition~(\ref{eq:independence}), 
\begin{eqnarray}
\left( \frac{n}{Mi} \right)^{\frac{1}{2}} \left(\frac{\check{\lambda}_i}{\lambda_i}
 -   1 \right) \overset{\rm D}{\longrightarrow} \mathcal{N}(0, 1)
\end{eqnarray}
holds under the following conditions
\begin{enumerate}
  \item $d \longrightarrow \infty$ and $n\longrightarrow  \infty$  for $i$ such that $\alpha_i  > 1/2$,
  \item $d \longrightarrow \infty$ and $d^{2-  4\alpha_i}/ndp) \longrightarrow 0$ for $i$ such that $\alpha_i  \in   (0, 1/2]$,
\end{enumerate}
where, superscript {\rm D} on the arrow stands for the convergence in distribution, and $\mathcal{N}(0, 1)$ is a standard Gaussian distribution. 
}
\end{comment}

Next, we apply the NR methodology to the PC direction vectors.
Define $\check{\vec{h}}_i \equiv (n\check{\lambda} _i)^{-1/2} \tilde{X} {\vec{u}}_i$, and examine its performance as the estimator of the PC direction vector $\vec{h}_i$. 
Note that $\check{\vec{h}}_i=(\hat{\lambda} _i/\check{\lambda} _i)^{-1/2}\hat{\vec{h}}_i$. 
%We assume $\check{\vec{h}}_i^\top \vec{h}_i \geq 0$ for all $i$ without a loss of generality. 
YA12\nocite{YATA2012193} showed the following theorem.

{\theorem{\bf [\citet{YATA2012193} III]}\label{th:YA2012d}\\
%{\theorem{\bf [\citet{YATA2012193} IV]}\label{th:YA2012d}\\
Assume $\lambda_i \; (i \leq m)$ has multiplicity one. 
Under conditions in Theorem~\ref{th:YA2012b}, and the components of $\tilde{Z}$ satisfy Condition~(\ref{eq:independence}), it holds that
\begin{eqnarray}
  \check{\vec{h}}_i^\top \vec{h}_i = 1 + o_{\rm P} (1).
\end{eqnarray}
}
We also consider the PC scores.  
The $i$-th PC score of $x_i$ is given by $\vec{h}_i^\top \vec{x}_j = \vec{z}_{ij} (\lambda_i)^{\frac{1}{2}} \equiv \vec{s}_{ij}$. 
Let ${\vec{u}}_i \equiv (\hat{u}_{i1}, \dots , \hat{u}_{in})^\top \; (i = 1, \dots , d)$. 
Consider we estimate the PC score by $\hat{\vec{u}}_{ij} (n\check{\lambda}_i)^{\frac{1}{2}} \equiv \check{\vec{s}}_{ij}$. 
We define the mean squared error (MSE) of the $i$-th PC score by 
\begin{eqnarray}
{\rm MSE}(\check{\vec{s}}_i) \equiv \frac{1}{n} \sum_{j=1}^n (\check{s}_{ij} - s_{ij})^2 \; . 
\end{eqnarray}
Then we have yet another important theorem.
{\theorem{\bf [\citet{YATA2012193} V]}\label{th:YA2012e}\\
Assume that $\lambda_i \; (i \leq m)$ has multiplicity one. 
Then, under the same conditions as Theorem~\ref{th:YA2012d}, 
\begin{eqnarray}
\frac{{\rm MSE}(\check{s}_i)}{\lambda_i} = o_{\rm P}(1) 
\end{eqnarray}
holds. 
}
\\

All these theorems hold for the cases with nonzero mean, by replacing the setting with zero-centered variables. 
Thus, now we have demonstrated that the NR methodology yields high performances for the estimation of the PC eigenvalues (contributions), PC vectors and PC scores for HDLSS data. 
In the following discussions, the presented eigenvalues are obtained by the NRPCA.

\subsubsection{Automatic Sparse PCA (A-SPCA)}

In this section, we introduce (thresholded) sparse PCA methods for high-dimensional data. 
The sparse PCA (SPCA) can be summarized as follows: 
Let $\hat{\vec{h}}_i=(\hat{h}_{i(1)},...,\hat{h}_{i(d)})^{\top}$ for all $i$. 
Given a sequence of threshold values $\zeta>0$, define the thresholded entries as 
\begin{align}
\hat{h}_{i*(s)}
=
\begin{cases}
\hat{h}_{i(s)} & \mbox{if $|\hat{h}_{i(s)}|\ge \zeta$}, \\
0 & \mbox{if $|\hat{h}_{i(s)}|< \zeta$} \label{ST1}
\end{cases} 
 \quad \mbox{ for $s=1,...,d$}.
\end{align} 
Let $\hat{\vec{h}}_{i*}=(\hat{h}_{i*(1)},...,\hat{h}_{i*(d)})^{\top}$. 
The thresholded estimator of ${\vec{h}}_{i}$ is defined by 
\begin{align}
\hat{\vec{h}}_{i**}
=\hat{h}_{i*}/\|\hat{h}_{i*}\|. 
 \label{ST2}
\end{align}
However, the estimator heavily depends on a choice of $\zeta$ and does not hold consistency properties for HDLSS data. 
Recently, \citet{YATA2022} proposed a SPCA method by using the NR method as follows: 
%%%%
For the PC direction by the NR method, 
$  \check{\vec{h}}_i=(\check{h}_{i(1)},...,\tilde{h}_{i(d)})^{\top}$, 
we assume $|\check{h}_{i(1)}|\ge \cdots \ge |\check{h}_{i(d)}|$ 
for the sake of simplicity. 
For a given constant $\omega\in (0,1]$, we consider PC directions whose cumulative contribution ratio is greater than or equal to $\omega$. 
There exists a unique integer $\check{k}_{i\omega} \in [1,d]$ such that 
\begin{align}
\sum_{s=1}^{\check{k}_{i\omega}-1}\check{h}_{i(s)}^2<\omega \ \mbox{ and } \ 
\sum_{s=1}^{\check{k}_{i\omega}}\check{h}_{i(s)}^2\ge \omega.
\label{SPCA}
\end{align}
We define that 
\begin{align}
\check{h}_{i \omega (s)}=
\begin{cases}
\check{h}_{i(s)} & \mbox{if $|\check{h}_{i(s)}|\ge  |\check{h}_{i(k_{i\omega})}|$
}, \\
0 & \mbox{otherwise} \notag
\end{cases} 
 \quad \mbox{ for $s=1,...,d$}.
\end{align}
Let $\check{\vec{h}}_{i\omega}=(\check{h}_{i \omega (1)},...,\check{h}_{i \omega (d)})^{\top}$.  
\citet{YATA2022} called the new PCA method that uses $\check{h}_{i \omega (s)}$s as the “automatic SPCA (A-SPCA)”. 
They showed some consistency properties of $\check{\vec{h}}_{i\omega}$ in the HDLSS data and investigated the performance in actual data analyses. 
Here the A-SPCA uses the same eigenvalues estimated by the NRPCA. 
In this paper, we set $\omega=1/2$ \footnote{{The value of  $\omega=1/2$ was obtained by simulations. }}.
Intuitively, the A-SPCA can extract the most important components of the eigenvectors obtained by the NRPCA. 
Namely, these components practically controls the PCs.

\section{Data} \label{sec:data}

Our target galaxy, NGC~253, is a nearby barred spiral galaxy at a distance of $3.5\;\mbox{Mpc}$, i.e., $1''$ corresponds to $17\; \mbox{pc}$ \citep{2005MNRAS.361..330R}. 
Morphological type Sc is assigned. 
It has a systemic heliocentric velocity of $243 \pm 2\; \mbox{km\,s}^{-1}$ ($z = 0.000811$: \citealt{2015MNRAS.452.3139K}), %\footnote{From NASA/IPAC Extragalactic Database (NED): \tt https://ned.ipac.caltech.edu/.}
mainly due to the cosmological expansion. 
This galaxy is one of the brightest sources of far infrared emission except the Magellanic Clouds and has been extensively studied at many wavelengths. 
It is known as a pure starburst, whose star formation rate (SFR) in the central molecular zone is estimated to be $\mbox{SFR} \sim 2 \; M_\odot \; [\mbox{yr}^{-1}]$ \citep{1980ApJ...238...24R,1999ApJ...518..183K,2015MNRAS.450L..80B}. 
Massive, dense, and warm ISM \citep[e.g.][]{2011ApJ...735...19S} were identified in the central region. 
Super star clusters are identified at optical and near-infrared (NIR) bands \citep{2009MNRAS.392L..16F}. 
An intense outflow is also observed \citep{2009ApJ...701.1636M,2013Natur.499..450B}. 
The super star clusters and outflow are also associated with the starburst activity. 

\subsection{ALMA map of NGC~253}

The data used in this work is a spectroscopic map of a nearby archetype starburst galaxy NGC~253, obtained by A17\nocite{2017ApJ...849...81A}, at Band~7. 
They obtained the 340--365~GHz ($\lambda \simeq 0.85$~mm) spectra covering a total frequency range of 11~GHz, compiling the results of two projects in the ALMA Cycle~2 observations: 2013.1.00735.S (PI: Nakanishi) and 2013.1.00099.S (PI: Mangum). 
The field of view of ALMA at these wavelengths is $16''\mbox{--}17''$. 
The observations cover the frequency range of 340.2--343.4~GHz and 352.5--355.7~GHz for 2013.1.00735.S, and 350.6--352.4~GHz and 362.2--365.2~GHz are covered by 2013.1.00099.S. 
The uncertainty of the flux calibration in ALMA Band~7 observations is $\sim 10$~\% according to the ALMA Cycle~2 Proposer’s Guide.
Data reduction is conducted with Common Astronomy Software Applications \citep[CASA:][]{2007ASPC..376..127M}, versions 4.2.1 and 4.2.2. 
The image of the central region of NGC~253 was created by using task {\tt clean}.
The synthesized beam size of the final images is $0.\hspace{-0.5mm}''45 \times 0.\hspace{-0.5mm}''3$, which corresponds to $8 \;\mbox{pc} \times 5\;\mbox{pc}$ at the distance to NGC~253. 
With a velocity resolution of $5.0\;\mbox{km\,s}^{-1}$, the rms noise levels are $1.0\mbox{--}2.4\; \mbox{mJy\,beam}^{-1}$. 
Various emission lines are detected in the spectra (see Fig.~2 Table~1 of A17\nocite{2017ApJ...849...81A}). 
The number of spectroscopic resolution unit (sampling number) is 2248. 

\begin{figure*}[tb]
    \centering
    \includegraphics[width=0.9\textwidth]{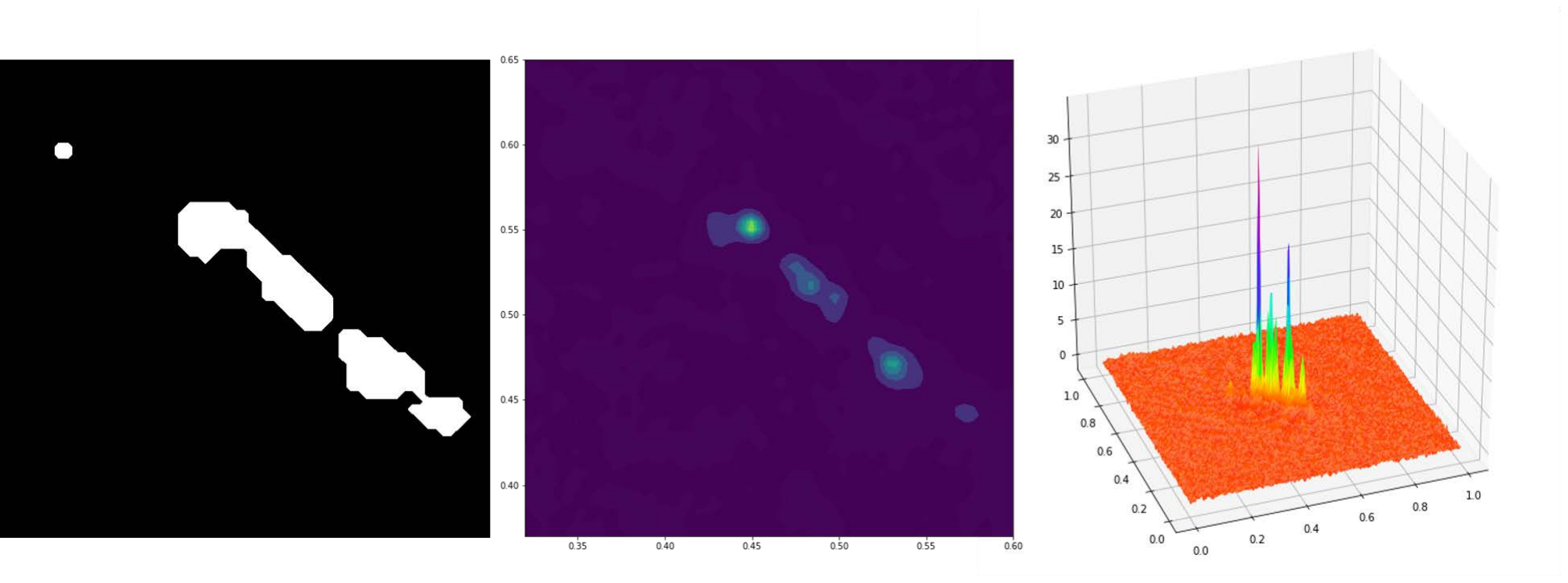}
    \caption{The bright regions of NGC~253 map cut out by the mask. 
    Left: the mask region map. White regions have significant intensity signals.
    Center: the cut-out region with significantly bright emission. 
    Right: the bird's view of the signal. 
    }\label{fig:NGC253_mask_map}
\end{figure*}

The A17\nocite{2017ApJ...849...81A} original map consists of $864 \times 864$ points in the spatial domain (R.A and dec.), and 2248 components along the frequency domain (in units of [Hz]) in total. 
However, since the pixel scale was smaller than the synthesized beam size of the map $0.\hspace{-0.5mm}''45 \times 0.\hspace{-0.5mm}''3$ (oversampled), each spatial grid in the map is not independent of each other. 
To overcome this issue, we regridded the map in the spatial dimension. 
The equivalent circular beam size corresponding to the elliptical beam of $0.\hspace{-0.5mm}''45 \times 0.\hspace{-0.5mm}''3$ is $0.\hspace{-0.5mm}''37$. 
We used this effective beam size to regrid the image map. 
The resulting map with independent grids consists of $230 \times 230$ pixels.
We then should cut out low-signal regions. 
For this, instead of estimating S/N for some lines, we integrated the intensity over the whole range of frequency and calculated the total intensity $I \; [\mbox{Jy\,beam}^{-1}]$ and chose the regions with {$I > 1.0$, based on the rms level.} 
We show this ``mask'' region, together with the significantly bright regions in Fig.~\ref{fig:NGC253_mask_map}.
The number of remaining pixels $n$ is 231.

\begin{figure}[tp]
    \centering
    \includegraphics[width=0.45\textwidth]{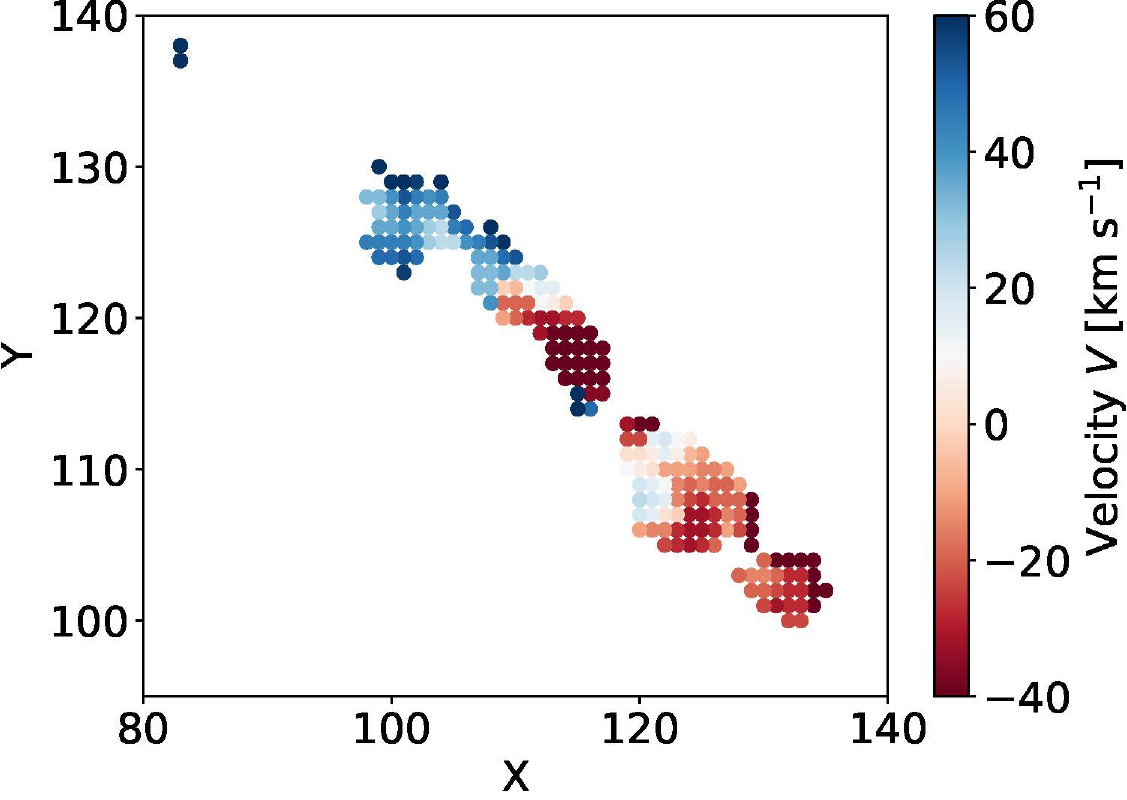}
    \includegraphics[width=0.45\textwidth]{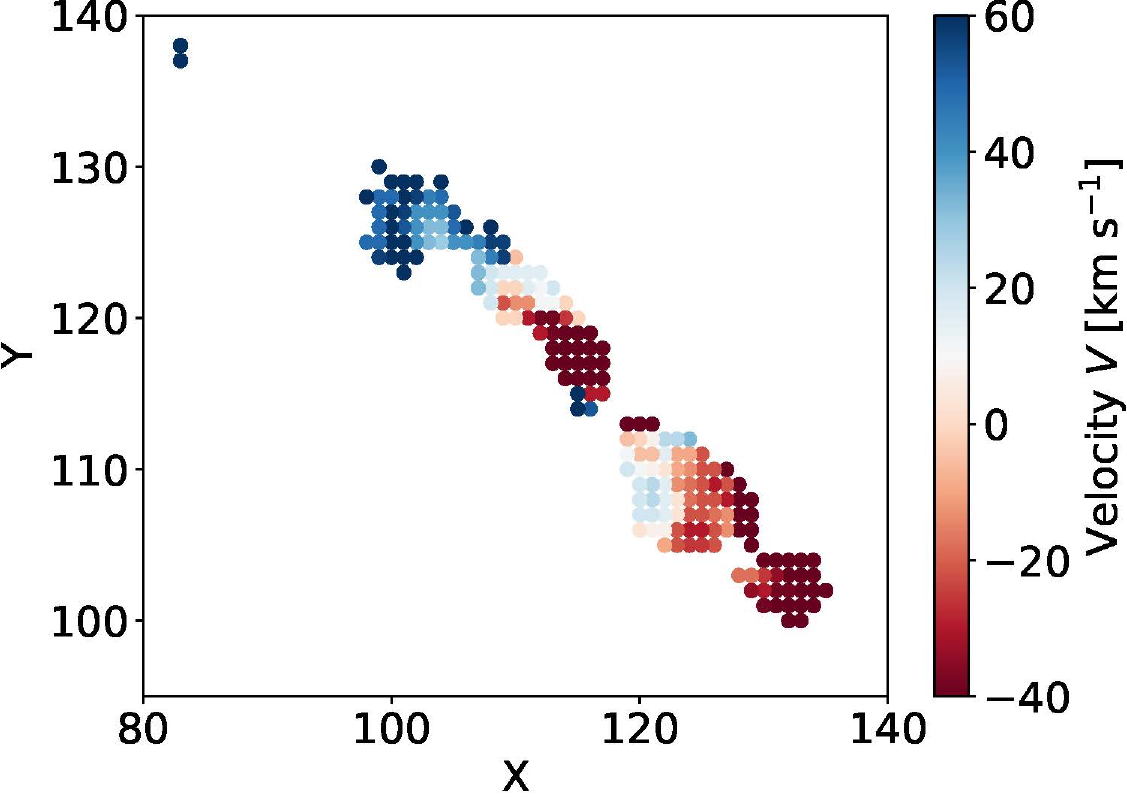}
    \includegraphics[width=0.45\textwidth]{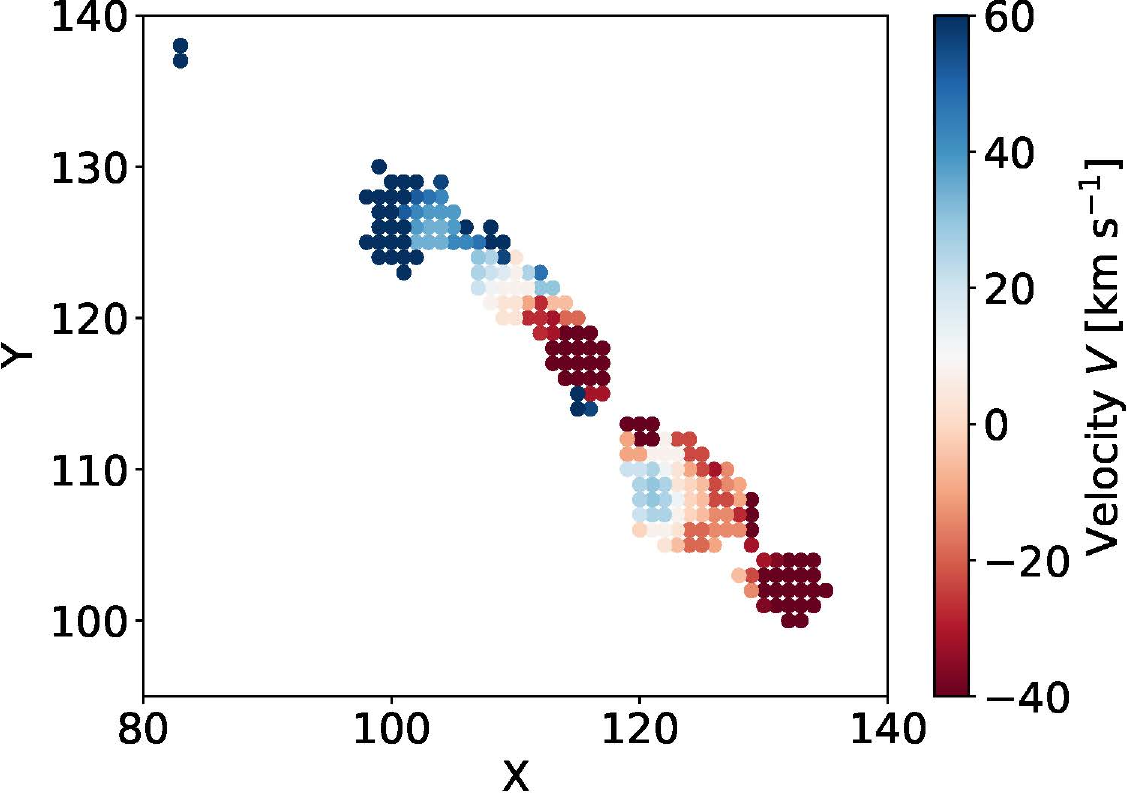}
    \caption{The velocity map of NGC~253 estimated from HCN(4--3) (Top), HNC(4--3) (Middle), and CS(7--6) (Bottom). 
    The systemic velocity of $243\;[\mbox{km\,s}^{-1}]$ is subtracted. 
    }\label{fig:NGC253_velocoty_corrected}
\end{figure}

\begin{figure}[t]
    \centering
    \includegraphics[width=0.85\textwidth]{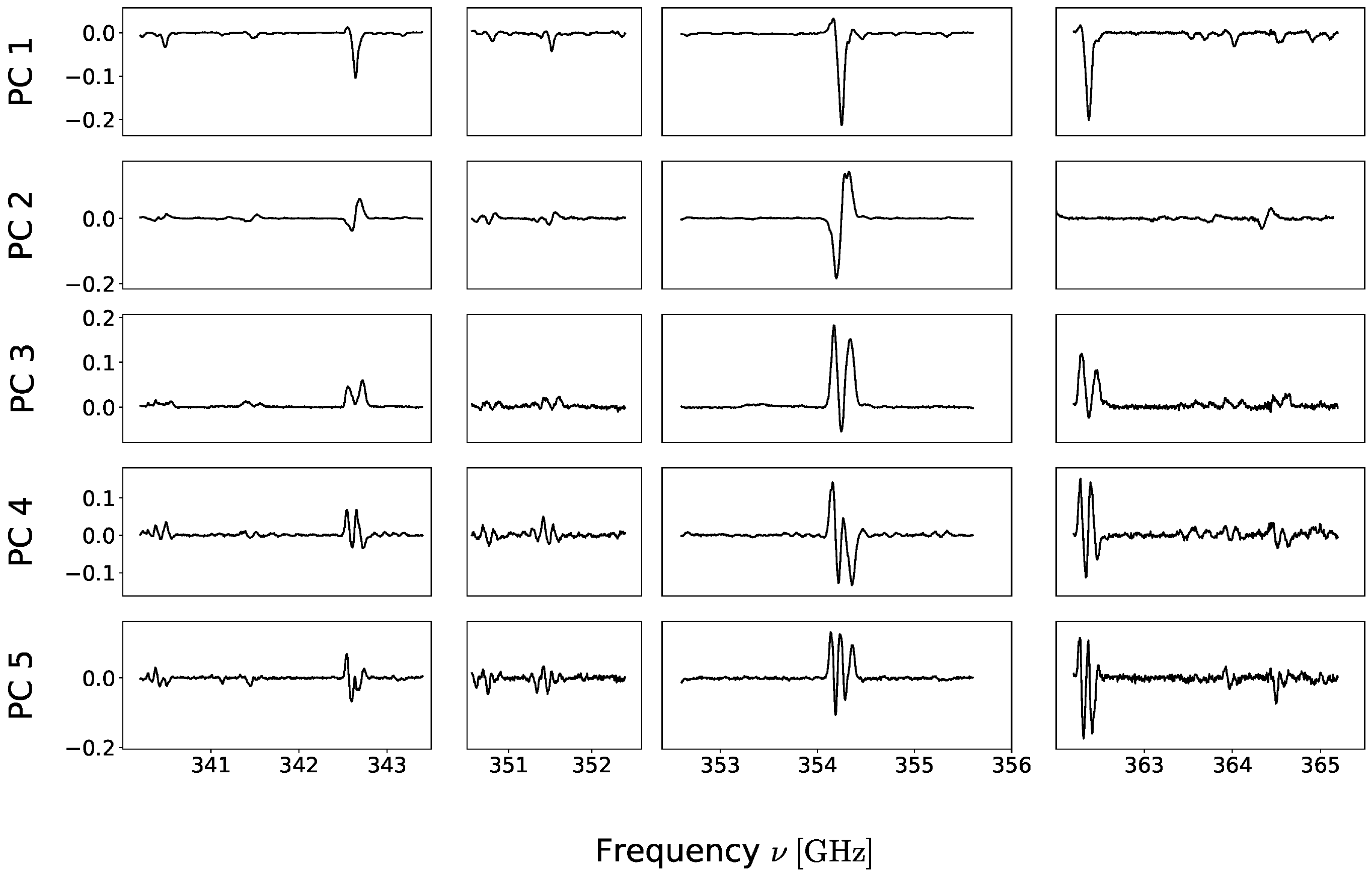}
    \includegraphics[width=0.85\textwidth]{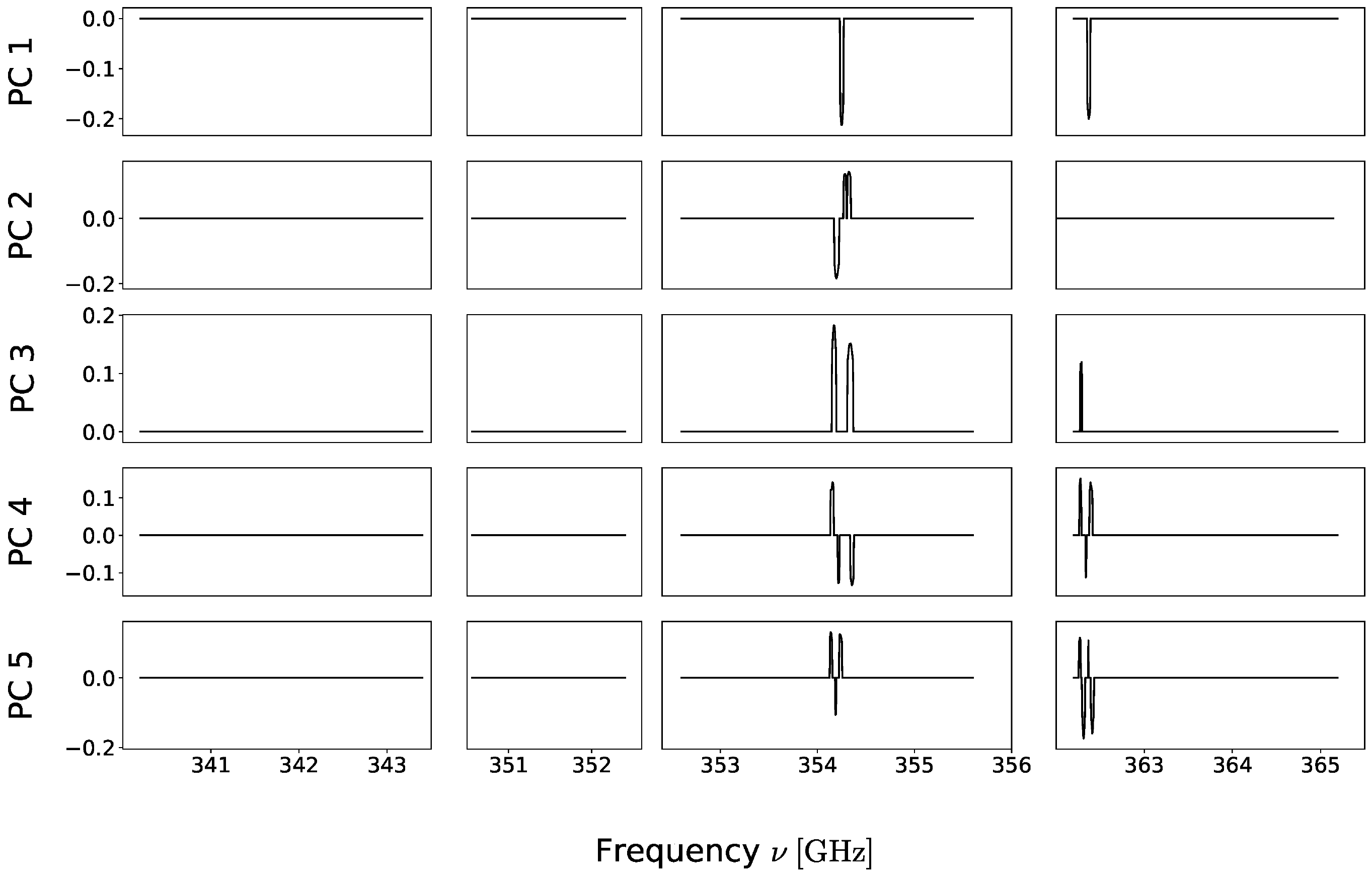}
    \caption{The eigenspectra corresponding to the 1st--5th principal components (PCs) constructed from the original ALMA spectral map of NGC~253. 
    {Upper five panels}: eigenspectra obtained by the noise-reduction principal component analysis (NRPCA).
    {Lower five panels}: same as Upper panels, but obtained by the automatic sparse principal component analysis (A-SPCA). 
    Only the controlling features have nonzero values.
    }
    \label{fig:eigenspectra_wo_doppler_correction}
\end{figure}

\subsection{Doppler shifts due to the systemic Hubble velocity and rotation}

The central region of NGC~253 is strongly inclined ($i \sim 78^\circ$). 
This causes a fairly coherent rotation pattern on the spectroscopic map through the Doppler shift. 
We made a Gaussian fit to three prominent emission lines HCN(4--3), HNC(4--3), and CS(7--6) in the spectra to estimate the line center wavelengths.
The map of the velocity field is presented in Fig.~\ref{fig:NGC253_velocoty_corrected}.
We determined the amount of the shift by averaging the shifts obtained from three lines. 
Then we shifted the frequencies with respect to the systemic velocity of NGC~253 caused by the Hubble velocity. 
After this correction, we have a map without the systematic effect of the coherent rotation and we can analyze more subtle properties of the submillimeter emission lines. 

We demonstrate an example of the emission lines in Fig.~\ref{fig:NGC253_shifted_emission} in Appendix~\ref{sec:Doppler_shift_correction}. 
We have frequency gaps in our ALMA data cube.
If we try to shift the spectrum according to the velocity, the boundaries of the spectra are also shifted. 
We simply adopted frequency ranges common to all the pixels. 
The final size of the data is $231$ along the spatial position, and $1971$ in spectral dispersion after the all corrections. 

\subsection{Spectral map as HDLSS data}

The resulting map of NGC~253 has $\mbox{spatial dimension } 231 \times \mbox{spectral dimension } 1971$, i.e., $n = 231$ and $d = 1971$, and clearly $n \ll d$.
Though it is not as extreme as DNA microarray data (typically $n \sim 10\mbox{--}100$ and $d \sim 10^4\mbox{--}10^5$), our ALMA map is typical HDLSS data \footnote{{
Note that $n > \log{d}=7.59$ when $d = 1971$ and $n=231$, 
so that the NR methodology works well for the HDLSS data. 
See Footnote 12.}}. 
Problems from astrophysical side is that a very large amount of information is contained in the spectra, and a large variety of spectral lines are observed compared to $n$. 
This situation is not special in the high-dimensional analysis. 

However, the application of the high-dimensional methods to spectroscopic data is not completely straightforward. 
Spectral lines are broadened and/or shifted by a turbulent motion and other physical mechanisms in the ISM.
Then, the information carried by each spectral resolution unit is not independent\footnote{
The existence of continuum radiation makes the situation even more complicated, but we do not try to examine this problem in the current work. }.
Such a complication has never been examined in other field of science about HDLSS data.
However, since it is commonly seen in spectroscopic data in any field of astronomy, before going to the detailed analysis, we first examine if the high-dimensional statistical analysis really works for the astronomical spectroscopic map, with line shifts and broadening. 

\begin{figure}[t]
    \centering
    \includegraphics[width=0.5\textwidth]{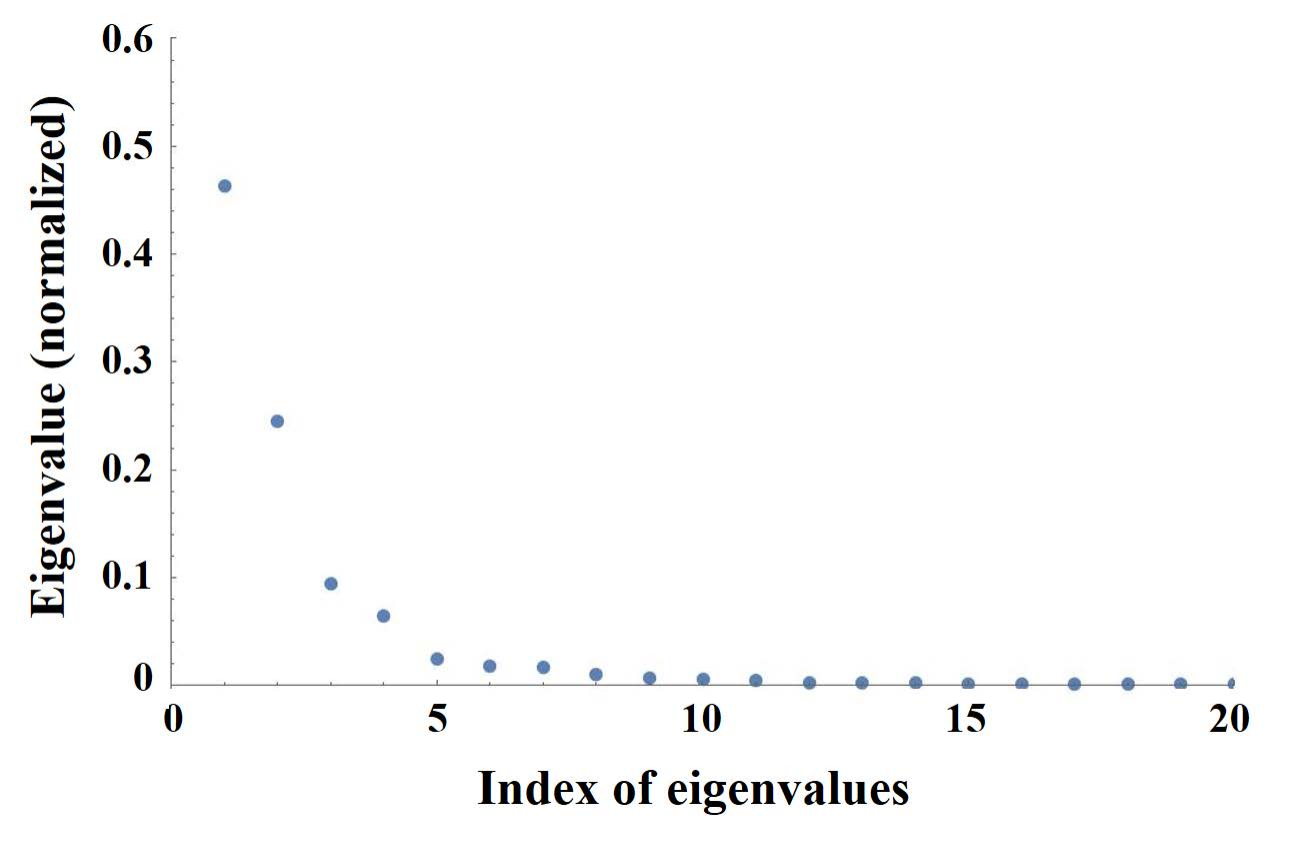}
    \caption{Normalized eigenvalues (contributions) obtained from the analysis of the ALMA spectroscopic map of NGC~253 by the NRPCA. 
    }\label{fig:NGC253_eigenvalue}
\end{figure}

\section{Preparatory Analysis}\label{sec:preparatory_analysis}

As we have already mentioned above, there are some potential problems specific to astronomical spectroscopic data, which are different from typical examples of HDLSS data in genomics and other fields. 
The most fundamental issue is that, when we regard a spectroscopic data as a vector data $\vec{x} = (x_1, \dots, x_d)^\top$, the information of neighboring frequency (or wavelength) index $i$ and $i+1 \; (i = 1, \dots, d-1)$ is not independent. 
Namely, we often have internal motion in the ISM of a galaxy that leads to a Doppler broadening along the frequency axis. 
We also find a systemic rotation of the system over the whole observed region of an object (\S~\ref{sec:data}). 
In order to examine if the high-dimensional statistical analysis would work for such type of astronomical spectroscopy data, we made a preparatory analysis of the original spectral mapping data of NGC~253 with the high-dimensional PCA.
By the term ``original'', we mean that we started this analysis without correcting the Doppler shift on the map due to the internal systemic rotation.
Namely, we blindly applied the high-dimensional PCA methods to the data and examined what feature would be detected.
For this, we sort the 2-dimensional coordinates along the R.A.- direction and then dec.\ direction, and renumbered so that we can regard the data as a 1-dimensional vector. 
Though we can apply the PCA directly to a 2-dimensional image, we adopted this procedure throughout this paper, as a first simplistic analysis.

\subsection{Eigenspectra}

In the following part of this work, we mainly focus on the A-SPCA results.
However, since the NRPCA deal with all the spectral features without extracting only the most important ones, we also present the eigenspectra (eigenvectors constructed from the observed spectra by PCA) from the NRPCA.
Upper panel of Fig.~\ref{fig:eigenspectra_wo_doppler_correction} shows the eigenspectra constructed by the NRPCA from the ALMA spectral map of NGC~253\footnote{
{
We evaluated the advantage of the NRPCA compared to the traditional PCA through the eigenvalue estimation. 
This is presented in Appendix~\ref{sec:comparison_PCA}.
}
}.
We observe that specific features are commonly found in similar frequency ranges of the spectra for all the PCs. 
This suggests that some spectral features contain richer information on the whole spectra than other spectral regions. 

The A-SPCA specifies and localizes the most important components that correspond to these particular features. 
Lower panel of Fig.~\ref{fig:eigenspectra_wo_doppler_correction} shows the eigenspectra obtained by the A-SPCA. 
As seen in Fig.~\ref{fig:eigenspectra_wo_doppler_correction}, the A-SPCA only leaves the values if the features are important to determine the spectra, and forces the other components to be zero. 
Hence, the A-SPCA eigenspectra are not used for the reconstruction of the spectra with all features, and the NRPCA is used for this purpose. 
We make use of this outstanding advantage of the A-SPCA to specify the controlling spectral features of the PCA in the following analysis.

\begin{figure*}[tb]
    \centering
    \includegraphics[width=0.8\textwidth]{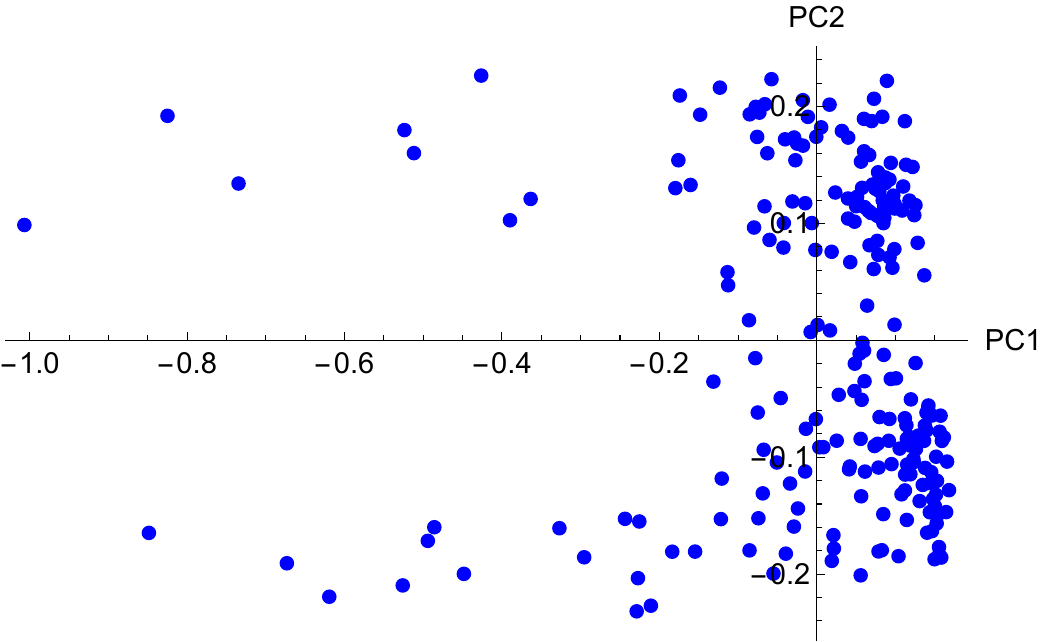}
    \caption{The distribution of PC1 and PC2 of the ALMA map of NGC~253 by the {A-SPCA}.    
%    Pairs of numbers associated with the symbols correspond to the spatial coordinates (angular positions) of the ALMA map. 
    }
    \label{fig:NGC253_PC}
\end{figure*}

\subsection{Distribution of the contribution of PCs}

First we show the eigenvalue distribution of the PCs sorted with their contribution obtained by the NRPCA in Fig.~\ref{fig:NGC253_eigenvalue}. 
First several eigenvalues are prominently larger than the rest. 
This behavior is referred to as ``spiked'' in the context of the high-dimensional statistical analysis. 
Recall that the eigenvalues of the PCA represent the contribution of each PC, namely, the PCA decomposes the whole covariance matrix into contributions from each PC.
Since the obtained eigenvalue distribution show{s} this spiked structure, we see that the high-dimensional PCA detected some characteristics of the ALMA spectral map of NGC~253. 
This guarantees that the current data can be safely regraded as typically HDLSS, and that the assumptions of the high-dimensional PCA apply.

Astrophysically, Fig.~\ref{fig:NGC253_eigenvalue} demonstrates a very promising possibility of the high-dimensional statistical analysis to the spectral mapping data. 
A17\nocite{2017ApJ...849...81A} identified more than 30 molecular/radical lines on the map (see their Fig.~1), as well as some broad or continuum features.
Even if we restrict our interest to the lines, it is implausible that we classify them only by physical intuition. 
Traditionally, for example, a line ratio has been used to extract physical information from spectral data. 
However, brute-force trials may not work if we try to diagnose many lines simultaneously, because the number of line ratios would increase prohibitively with a number of features, known as a combinatorial explosion. 
In a clear contrast, Fig.~\ref{fig:NGC253_eigenvalue} implies that the whole information of the complicated spectral map can be represented by first several PCs. 
This means that the complicated spectral features including more than 30 lines are controlled by several bases represented by the PCs. 
This advantage itself is a well-known benefit to use the PCA in general as a dimensionality reduction method \citep[e.g.,][among many others]{1998A&A...332..459G,1999MNRAS.303..284R,2011MNRAS.411.1809W,2019ApJ...883...82P,2020AJ....160...45P}, but all these previous applications were on integrated 1-d spectra with sufficiently large number of $n$, namely, not HDLSS data. 
We stress that the analysis in this work is substantially different from these works: the current method is able to be applied to a spectral map which is a typical HDLSS dataset and traditional PCA would not have worked due to the huge noise sphere, as mentioned in Section~\ref{sec:high_dimension}. 

Since we confirmed that the methodology of high-dimensional statistical analysis worked well for the astrophysical spectral amp data, we safely proceed to further analysis by the high-dimensional PCA.

\subsection{Physical meaning of first two PCs}

\begin{figure}[tb]
    \centering
    \includegraphics[width=0.45\textwidth]{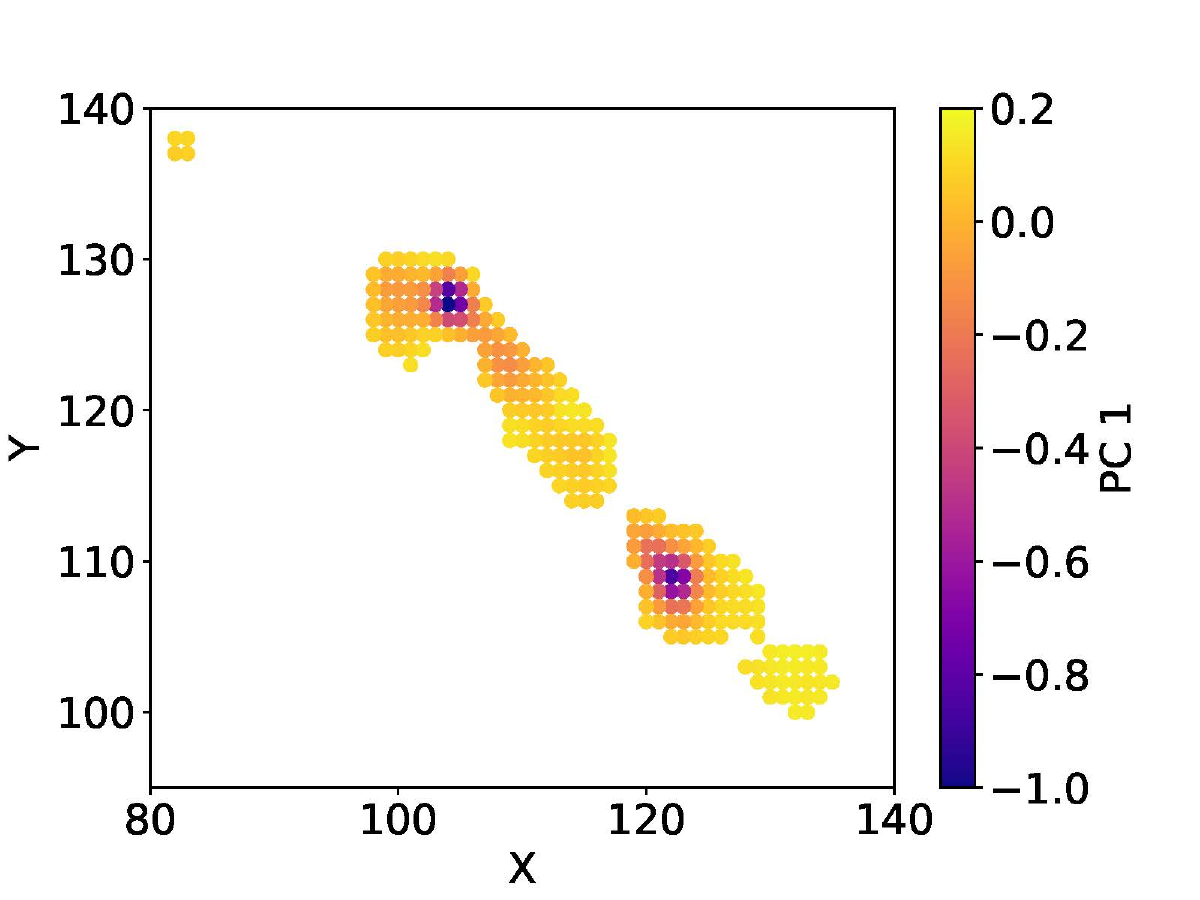}
    \includegraphics[width=0.45\textwidth]{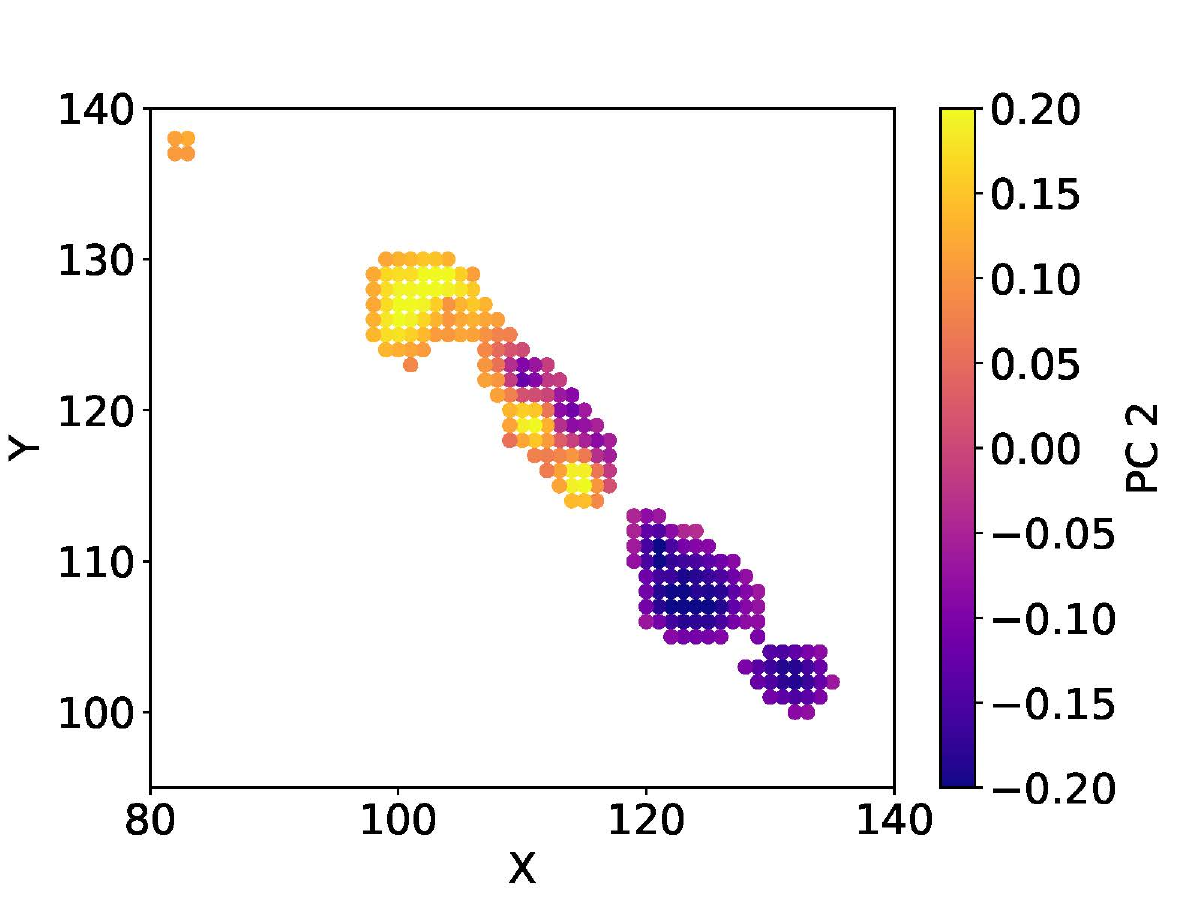}
    \caption{The 2-dimensional structure to map PC1 and PC2 obtained by the A-SPCA.
    Left panel shows the map of PC1, and Right panel describes the map of PC2, respectively. }
    \label{fig:NGC253_map_PC}
\end{figure}

\begin{figure*}[tb]
    \centering
     \includegraphics[width=0.4\textwidth]{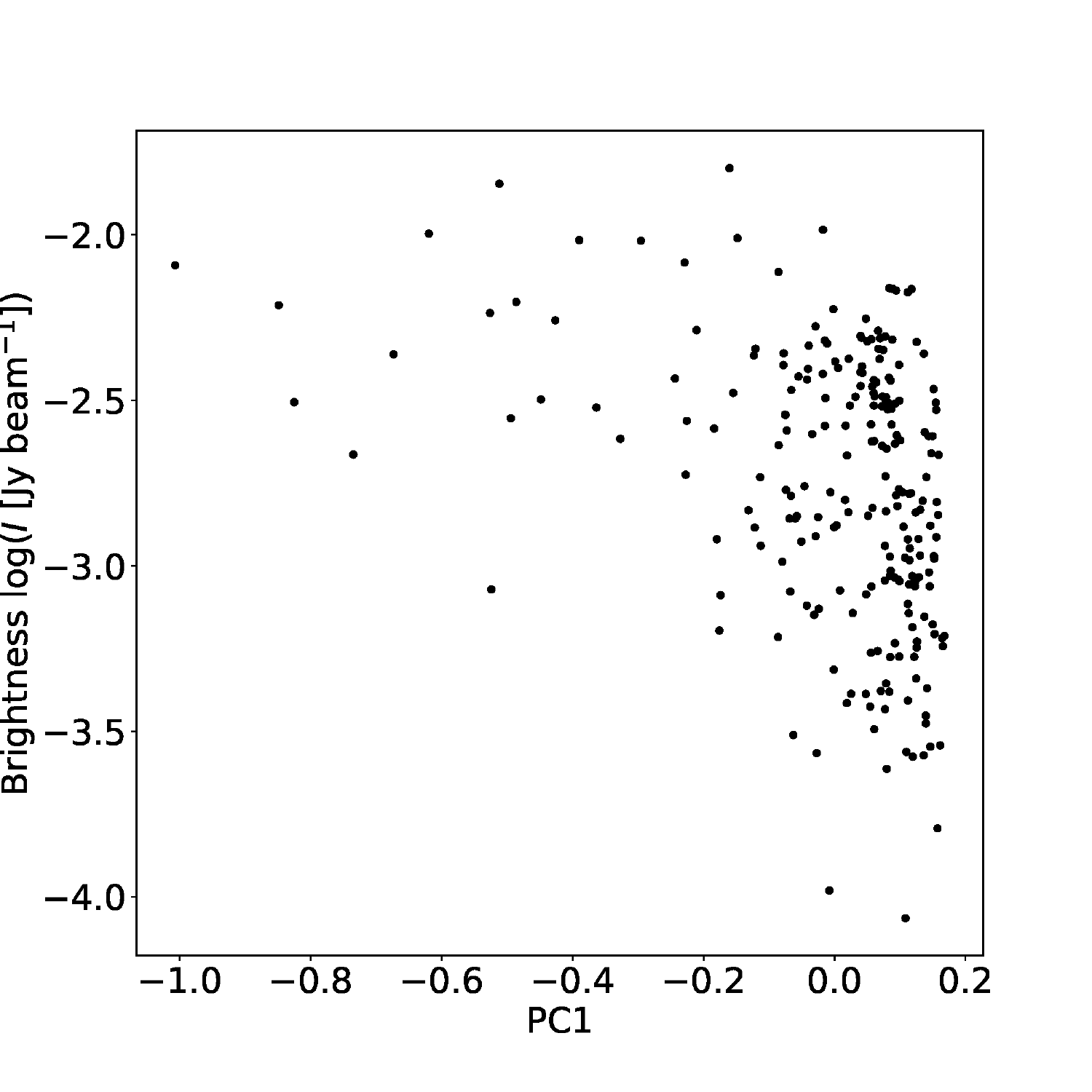}
    \centering
     \includegraphics[width=0.4\textwidth]{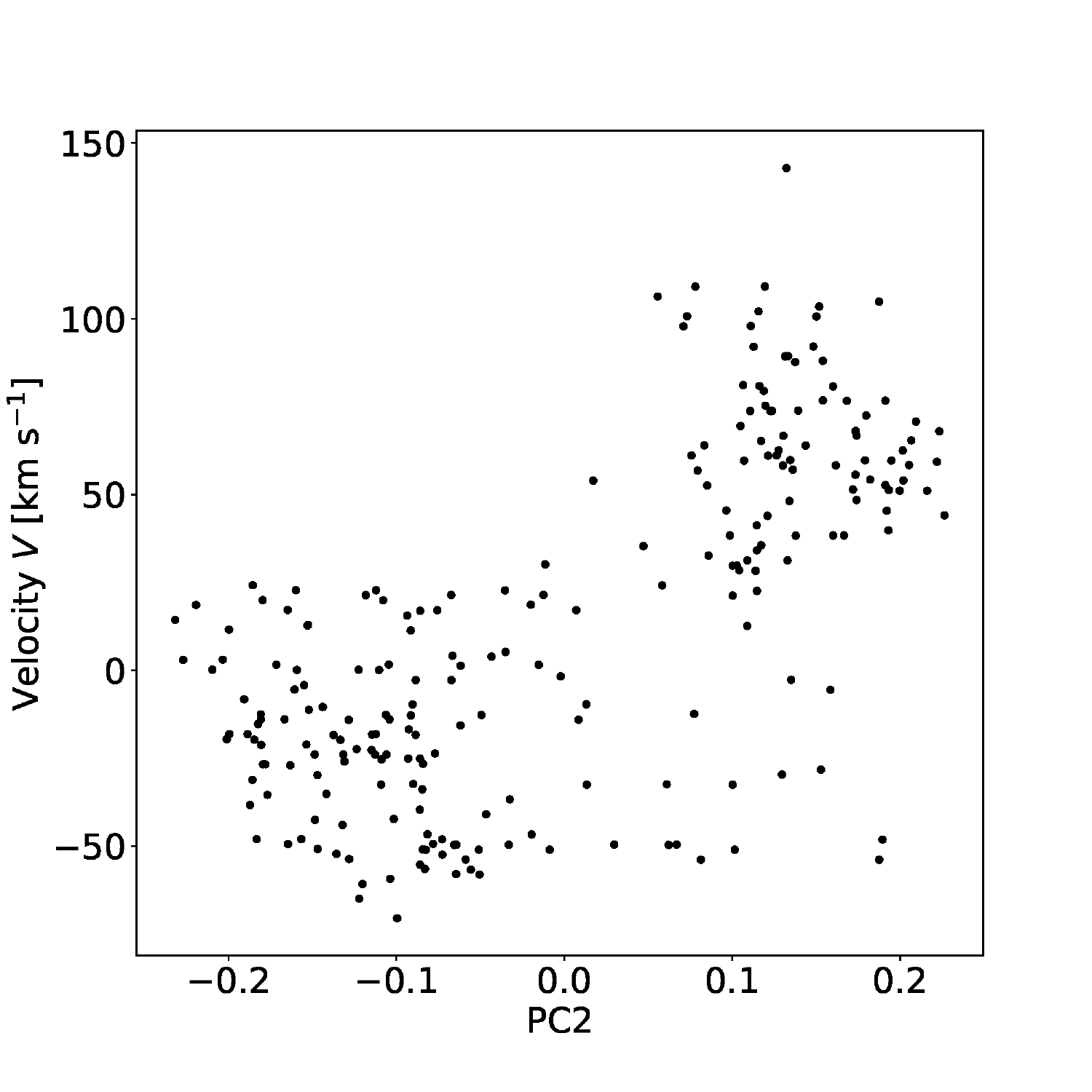}
    \caption{The scatter plot between PCs and observed physical properties. 
    Left: the correlation between PC1 and the integrated intensity over the whole frequency range $I \; [\mbox{Jy\,beam}^{-1}]$ on the ALMA map of NGC~253.
    Right the correlation between PC2 and the peculiar velocity on the ALMA map.}
    \label{fig:NGC253_scatter_PCs}
\end{figure*}

Next step is to examine what PC1 and 2 represent. 
Figure~\ref{fig:NGC253_PC} shows the scatter of PC1 and PC2.
%A pair of numbers associated with each symbol represents the indices of coordinates on the ALMA map. 
The most striking feature is the butterfly-like pattern symmetric with respect to the $x$-axis. 
It reminds us of the coherent rotation of the central region of NGC~253. 
We reconstructed the 2-dimensional structure to map PC1 and 2 in Fig.~\ref{fig:NGC253_map_PC}. 

It is clear that PC1 represents the intensity of lines, and PC2 represents the symmetric and coherent pattern of the center of NGC~253. 
We observe a nice agreement between Lower panel of Fig.~\ref{fig:NGC253_map_PC} with directly estimated velocity field of the same region in Fig.~\ref{fig:NGC253_velocoty_corrected}.

{}To have a deeper look at the relation, we present the correlation between PCs and observed properties in Fig.~\ref{fig:NGC253_scatter_PCs}. 
Upper panel of Fig.~\ref{fig:NGC253_scatter_PCs} is the scatter plot between PC1 and the total intensity integrated over the whole frequency range for each pixel. 
We indeed observe a good correlation between them (Fig.~\ref{fig:NGC253_scatter_PCs}). 
The correlation is not linear, because the PCA extracts only the linear features of the data, while the relation to the physical quantities would not be necessarily linear. 
We also examined the relation between the peculiar velocity and PC2 in {Right panel} of Fig.~\ref{fig:NGC253_scatter_PCs}.
As we already presented in 2-dimensional map, PC2 approximately delineate the velocity field in Fig.~\ref{fig:NGC253_scatter_PCs}.
However, we see some outliers where the PC2 and the peculiar velocity are not coherent.
These regions may be affected by some local phenomenon, for example, a strong outflow. 
We will revisit this issue later with a detailed analysis.

\subsection{Spectral features corresponding to PCs: original data}\label{sec:features_PCs}

\begin{figure}[tbh]
    \centering\includegraphics[width=0.9\textwidth]{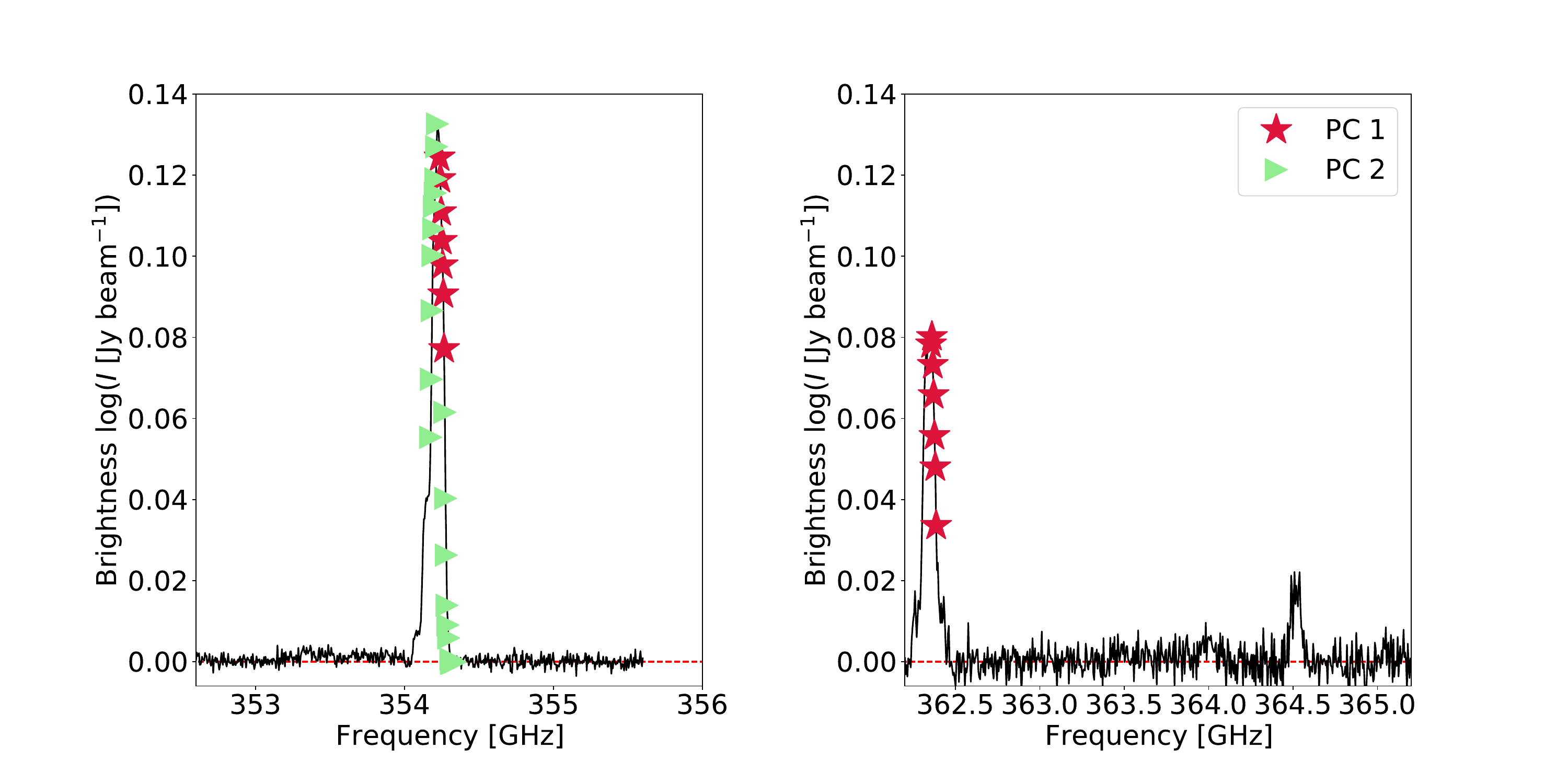}
    \caption{Responsible features to characterize PCs from the A-SPCA for the ALMA map of NGC~253. 
    Stars represent the features responsible for PC1, and triangles are for PC2.
    }
    \label{fig:features_original}
\end{figure}

It is interesting to examine what features are concretely responsible for PCs. 
Actually the A-SPCA can specify the responsible features to characterize the PCs.
We show the responsible spectral features in Figs.~\ref{fig:features_original}. 
The spectra are integrated over the whole analyzed regions for both figures.
In Fig.~\ref{fig:features_original}, stars represent the features responsible for PC1, and triangles are for PC2.
We do not show higher-order contributions for clarity. 
PC1-related features are assigned to the HCN(4--3) and HNC(4--3) lines, particularly close to the central part of the lines. 
PC2-related features are related to HCN(4--3), but in contrast to PC1, they are connected to the wing part of the line. 
Since the spectra are integrated over the analyzed regions, the wing is made actually by the Doppler shifts. 
This is clearly consistent with the fact that PC2 map (Fig.~\ref{fig:NGC253_map_PC}) agrees well with the Doppler velocity field of NGC~253 map (Fig.~\ref{fig:NGC253_velocoty_corrected}).

%Thus, we demonstrated the power of the high-dimensional PCA to extract the features of the emission lines on the ALMA map.  
The performance of the A-SPCA to the ALMA spectroscopic map as HDLSS data is certainly guaranteed by this preparatory analysis. 
We can safely proceed to the detailed analysis of the molecular lines, as in the next Section. 

\begin{figure}[t]
    \centering
    \includegraphics[width=0.85\textwidth]{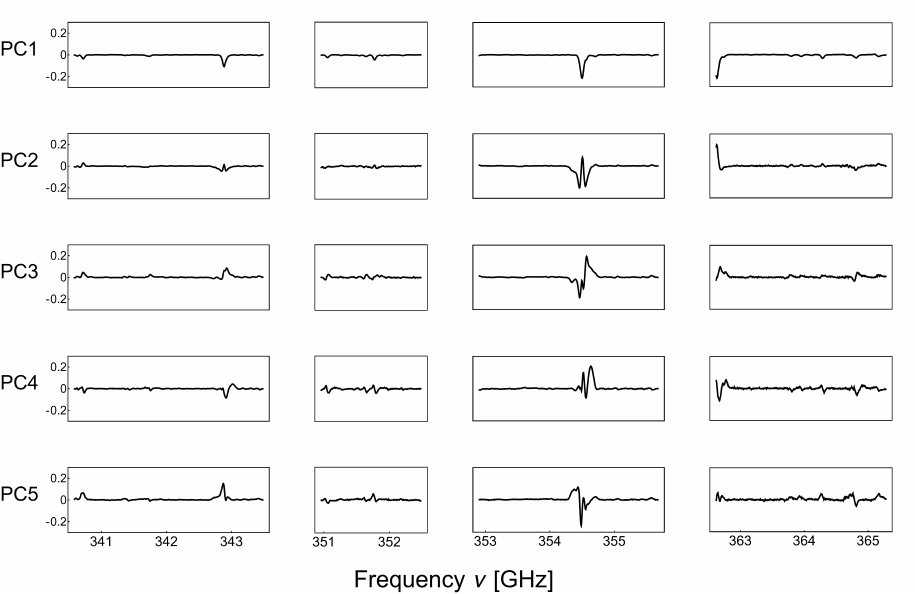}
    \includegraphics[width=0.85\textwidth]{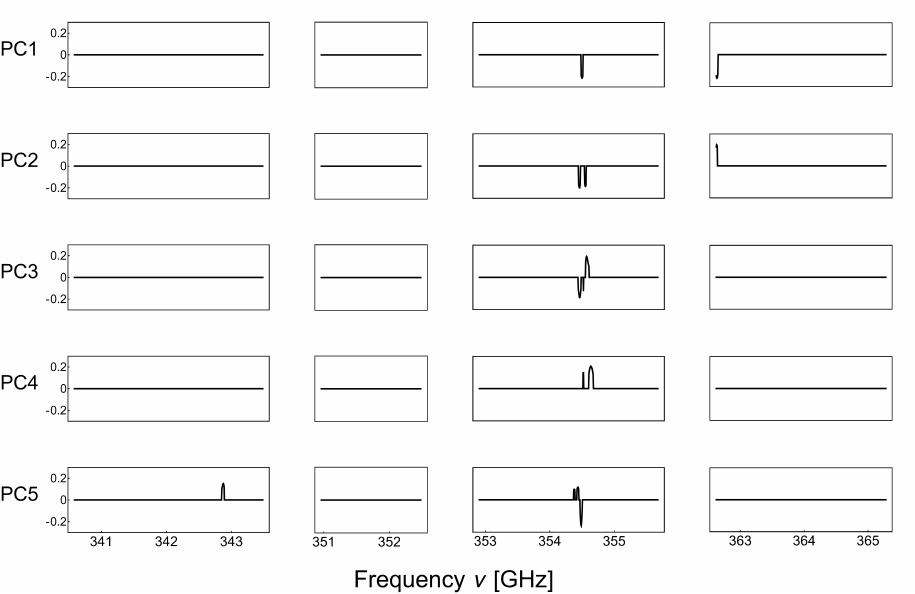}
    \caption{The eigenspectra corresponding to the 1st--5th principal components (PCs) constructed from the Doppler-corrected ALMA spectral map of NGC~253. 
    {Upper five panels} presents the eigenspectra obtained by the NRPCA, and {Lower five panels} shows that obtained by the A-SPCA.}
    \label{fig:eigenspectra_w_doppler_correction}
\end{figure}

\begin{figure}[t]
    \centering
    \includegraphics[width=0.85\textwidth]{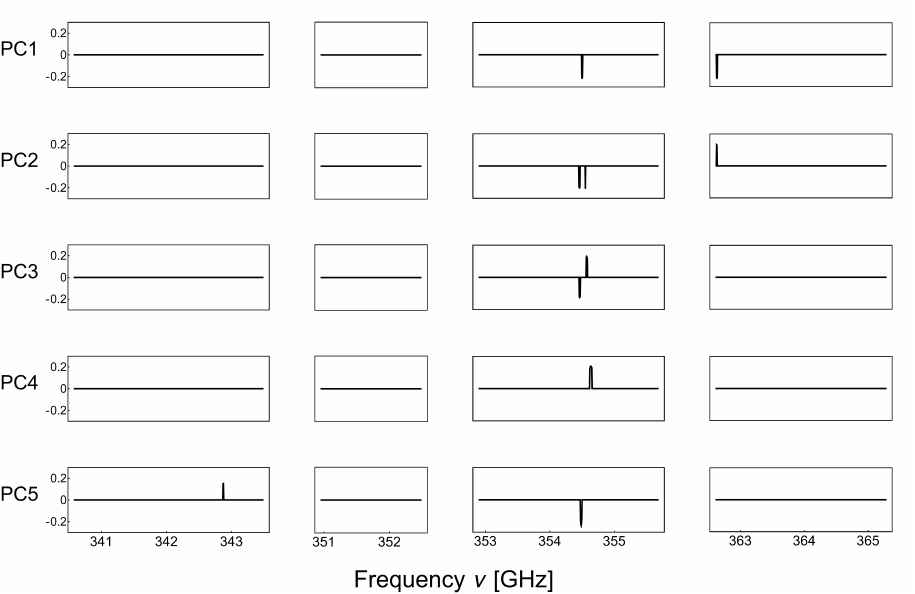}
    \includegraphics[width=0.85\textwidth]{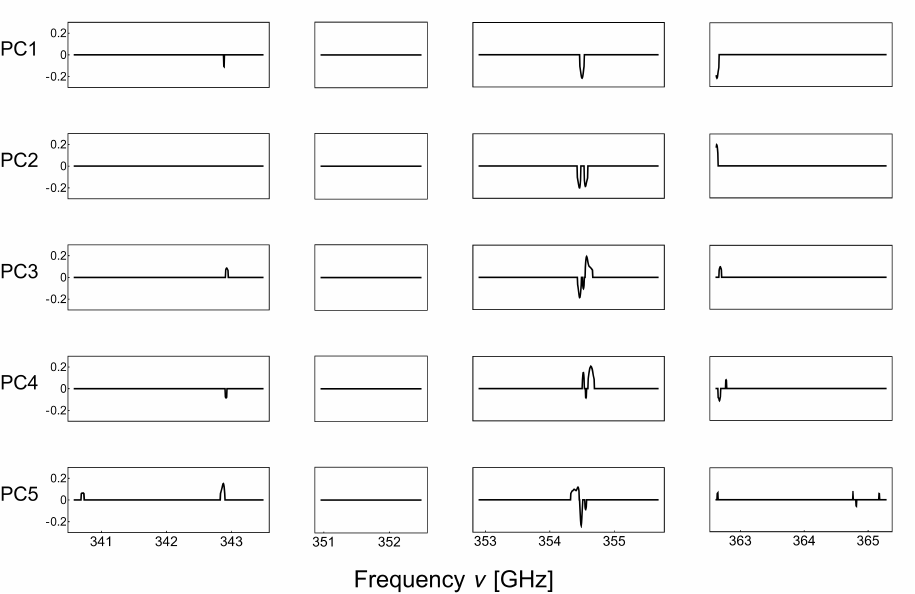}
    \caption{{The eigenspectra corresponding to the 1st--5th principal components (PCs) constructed from the Doppler-corrected ALMA spectral map of NGC~253. 
    {Upper five panels} presents the eigenspectra obtained by the A-SPCA with 0.25, and {Lower five panels} shows that obtained by the A-SPCA with 0.75.}}
    \label{fig:eigenspectra_w_doppler_correction_0.25_0.75}
\end{figure}

\section{Results and Discussion}\label{sec:results_discussion}

After examining the performance of the high-dimensional statistical analysis in \S~\ref{sec:preparatory_analysis}, now we can proceed to a further physical analysis. 
\subsection{Application of the high-dimensional PCA to the Doppler shift-corrected map of NGC~253}

We apply the high-dimensional PCA to the Doppler-corrected map of NGC~253, exactly in the same manner as we did in \S~\ref{sec:preparatory_analysis}. 
Note that, though we use the same notation for PC2, now PC2 is the value obtained from the Doppler-corrected map and different from those in \S~\ref{sec:preparatory_analysis} from the NRPCA. 
We should also note that since we applied the NRPCA and A-SPCA to the Doppler-corrected map, all the contributions of PCs are affected by the correction.

Same as the preparatory analysis, we first show the eigenspectra of the Doppler-corrected spectral map in Fig.~\ref{fig:eigenspectra_w_doppler_correction}. 
Comparing with Fig.~\ref{fig:eigenspectra_wo_doppler_correction}, the meaning of the shape of each eigenspectrum becomes clearer. 
We stress that since we have already subtracted the global shift of the whole spectra, {\sl it does not represent the systemic velocity of the position anymore}. 
Thus, the velocity structure is always the first-order deviation from the systemic rotation pattern, and appears as a broadening or asymmetric distortion of the line profiles. 
Another important difference is that an important spectral line is identified in PC5 (see PC5 of Bottom panel in Fig.~\ref{fig:eigenspectra_w_doppler_correction}) which was not specified from the original data. 
This is because the dominant effect of the systemic rotation was successfully removed from the map, and more subtle features appear more prominent in the Doppler-corrected map. 
Now very clearly PC2 represents the wing of the HCN and some other lines, indicating that it reflects a very local expansion of the gas.
PC3 and 4 describe the blueward distortion of the lines, caused by a gas motion toward us faster than the expansion velocity. 
Though PC5 is less prominent, but similarly to PC3 and 4, it represents the redward distortion of the lines. 

\begin{figure*}[tb]
    \centering
    \includegraphics[width=0.5\textwidth]{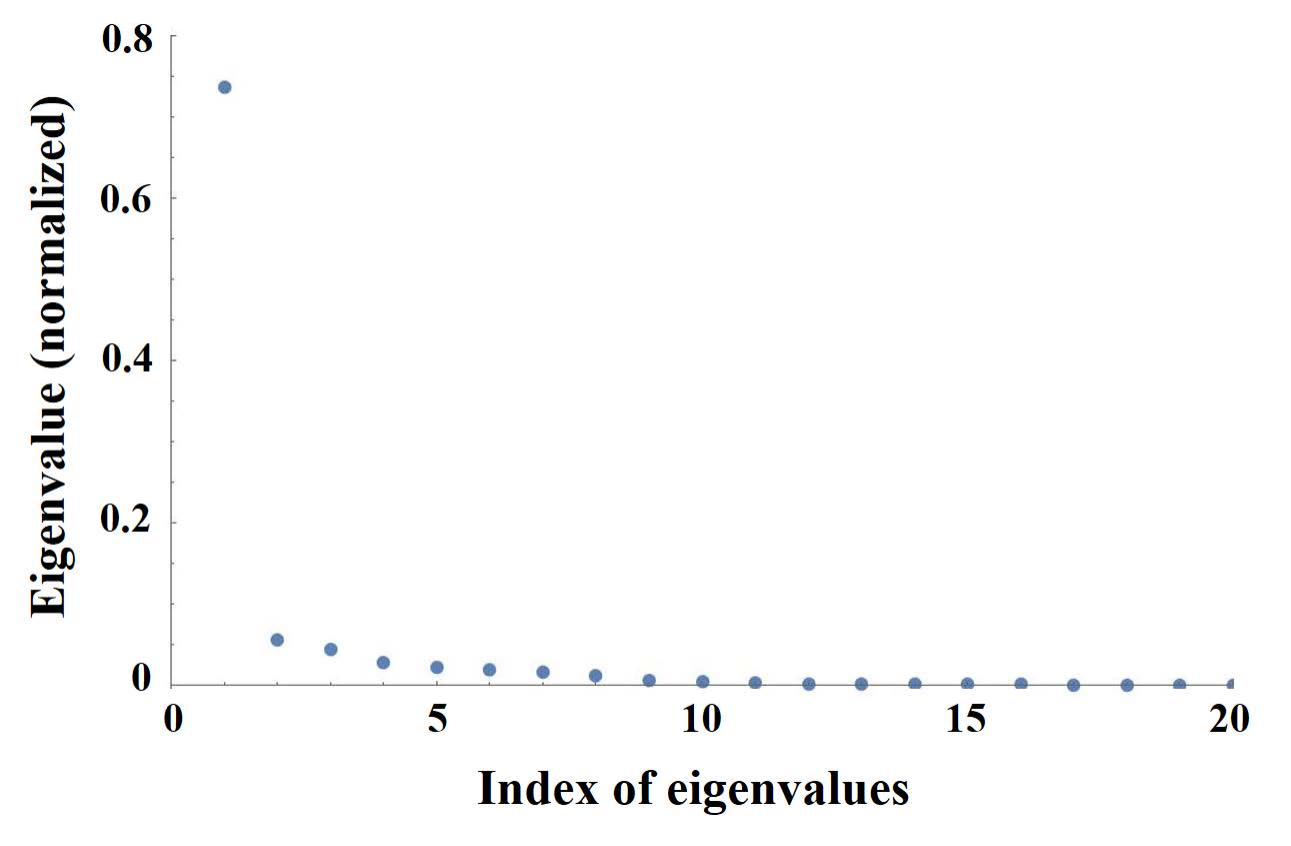}
    \caption{Normalized eigenvalues (contributions) obtained from the analysis of the ALMA spectroscopic map of NGC~253 by the NRPCA, after correcting the effect of the Doppler shifts of the systemic rotation. }
    \label{fig:NGC253_eigenvalue_Doppler_corrected}
\end{figure*}

We present the distribution of the contribution of PCs in Fig.~\ref{fig:NGC253_eigenvalue_Doppler_corrected}. 
Now the contribution of new PC2 is much smaller than old PC2 in Fig.~\ref{fig:NGC253_eigenvalue}. 
This is also because of the successful removal of the the systemic rotation. 
Instead, we have a smaller but non-negligible contribution of new PC2. 
Now PC2 would describe much more subtle and local features of the spectral line map of NGC~253, which is consistent with the implications from the eigenspectra. 

\begin{figure*}[p]
    \centering\includegraphics[width=0.8\textwidth]{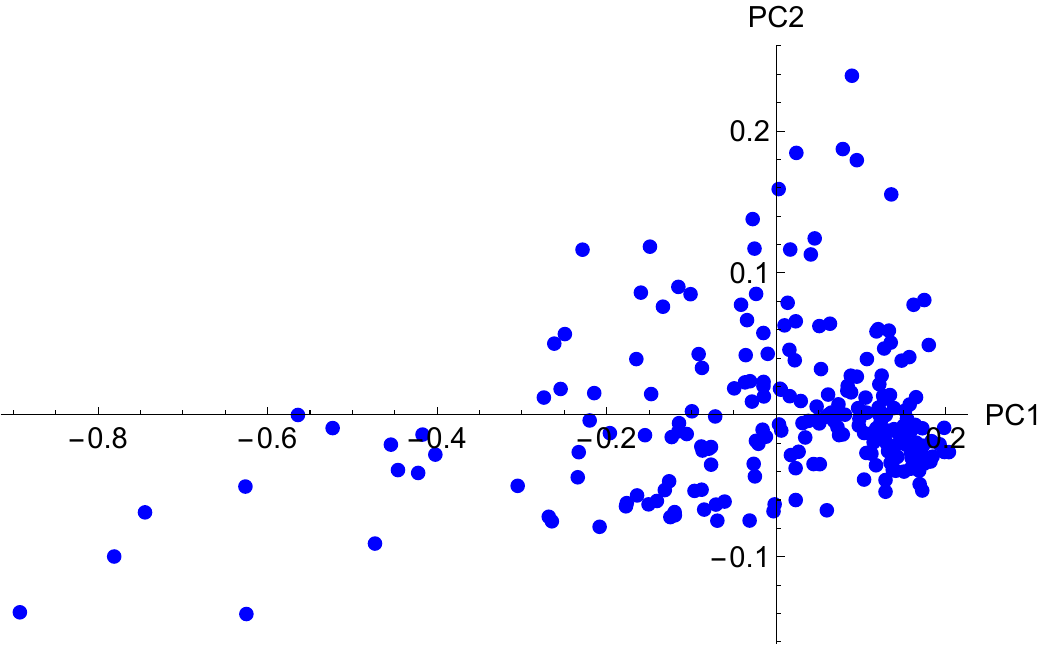}
    \caption{The 2-dimensional structures of the distribution of PC1 and PC2 of the NGC~253 obtained by the {A-SPCA}, after correcting the effect of the Doppler shifts of the systemic 
    Bottom: same as Top panel, but with PC1 and PC3. 
    Note that the PC2 in this figure is different from that in Fig.~\ref{fig:NGC253_map_PC}.
    }
    \label{fig:NGC253_PC_Doppler_corrected_12}
    \centering\includegraphics[width=0.8\textwidth]{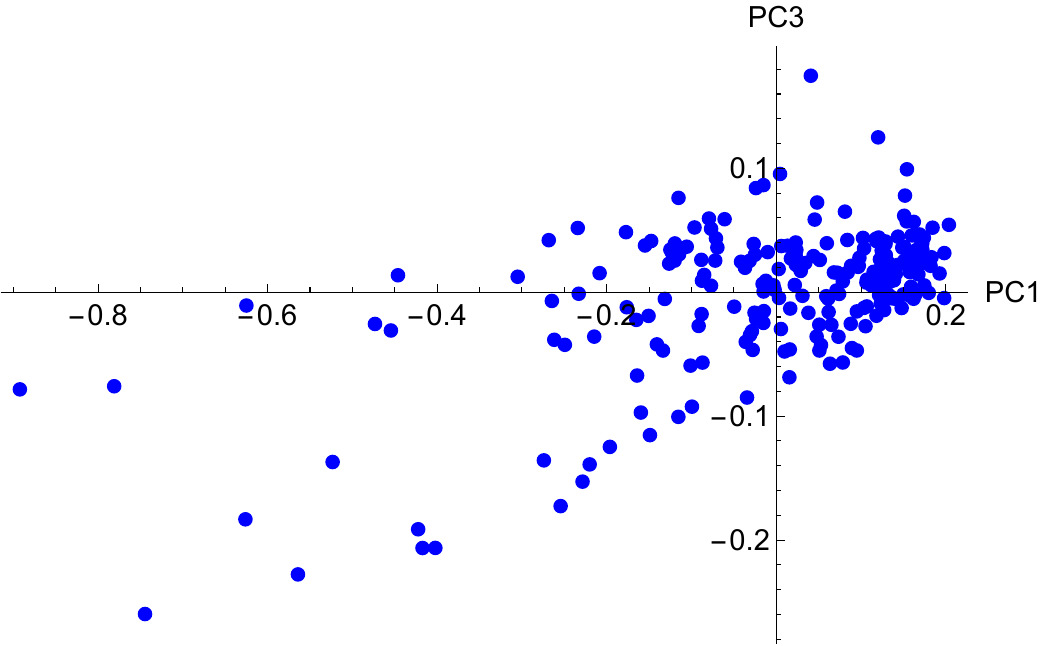}
    \caption{Same as Fig.~\ref{fig:NGC253_PC_Doppler_corrected_12}, but for PC1 and PC3. 
    }
    \label{fig:NGC253_PC_Doppler_corrected_13}
\end{figure*}

\begin{figure*}[t]
    \centering\includegraphics[width=0.8\textwidth]{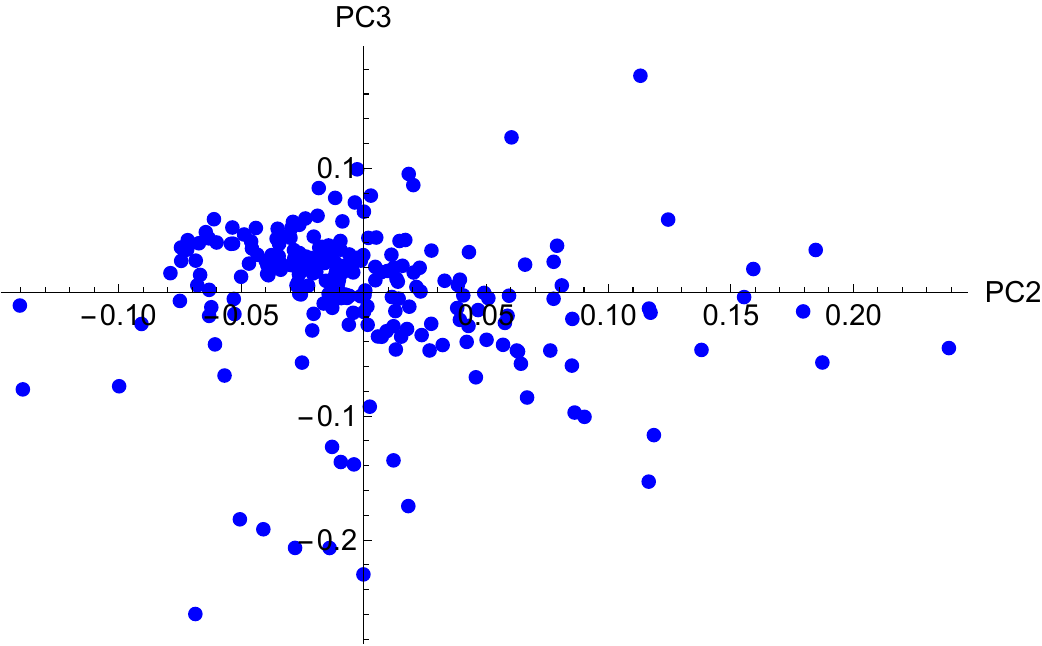}
    \caption{Same as Fig.~\ref{fig:NGC253_PC_Doppler_corrected_12}, but for PC2 and PC3. 
    }\label{fig:NGC253_PC_Doppler_corrected_23}
\end{figure*}

This is more drastically represented on the scatter plot of PCs. 
We demonstrate the relations between PCs in Figs.~\ref{fig:NGC253_PC_Doppler_corrected_12}--\ref{fig:NGC253_PC_Doppler_corrected_23}. 
Figure~\ref{fig:NGC253_PC_Doppler_corrected_12} shows the 2-dimensional distribution of PC1 and PC2. 
The butterfly-like pattern seen in Fig.~\ref{fig:NGC253_PC} completely disappeared. 
Figures~\ref{fig:NGC253_PC_Doppler_corrected_13} and \ref{fig:NGC253_PC_Doppler_corrected_23} are the relations between PC1 and PC3, and PC2 and PC3, respectively. 
Again we do not find any symmetric pattern in Figs.~\ref{fig:NGC253_PC_Doppler_corrected_13} and \ref{fig:NGC253_PC_Doppler_corrected_23}. 
%A bird's view of the distribution of PC1, 2, and 3 is presented in Fig.~\ref{fig:NGC253_PC_Doppler_corrected_123}.
{}From these figures, we do not have an obvious symmetric pattern in the intrinsic spectral features in the map. 
Both in 2- and 3-d, we observe a continuous distributions among PCs.  

\subsection{Map of PCs of the NGC~253 spectra}
\begin{figure*}[t]
    \centering
    \includegraphics[width=0.45\textwidth]{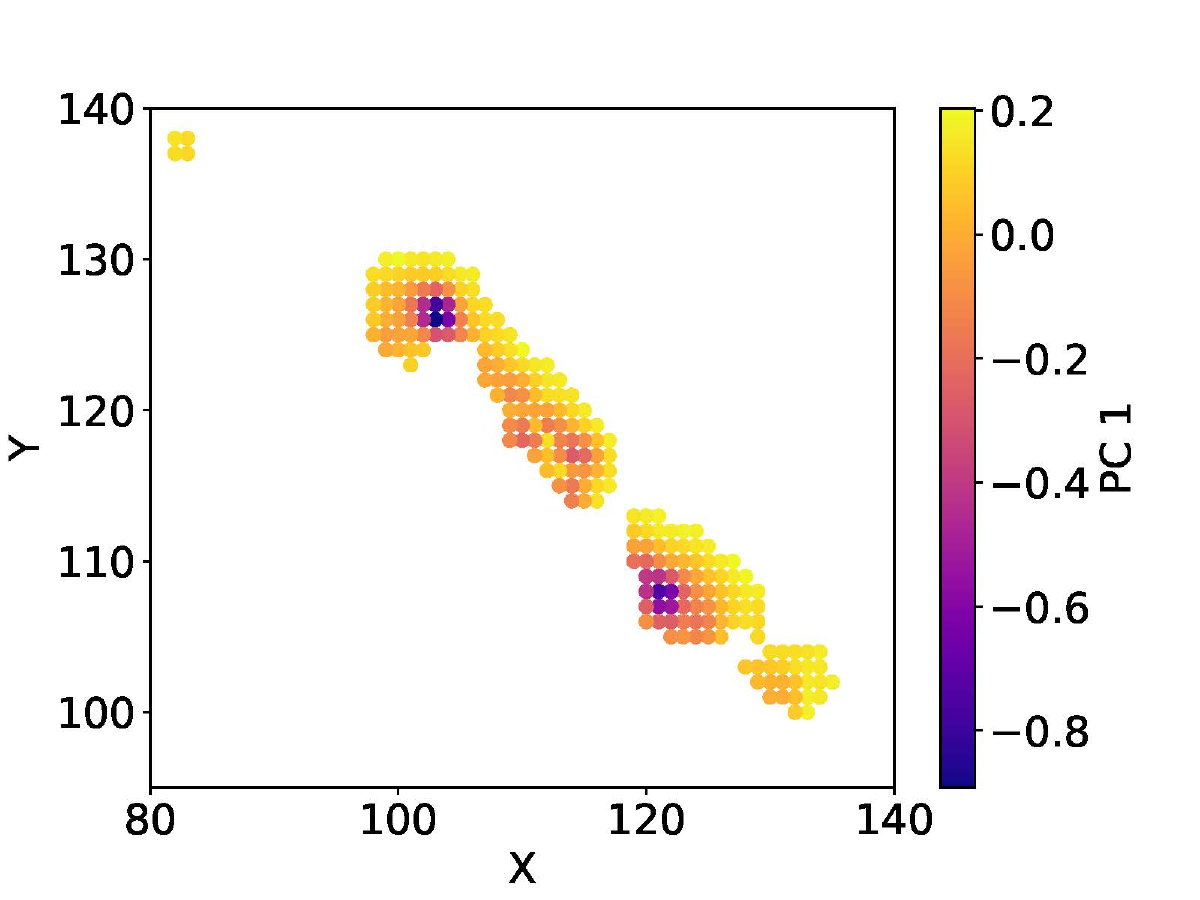}
    \includegraphics[width=0.45\textwidth]{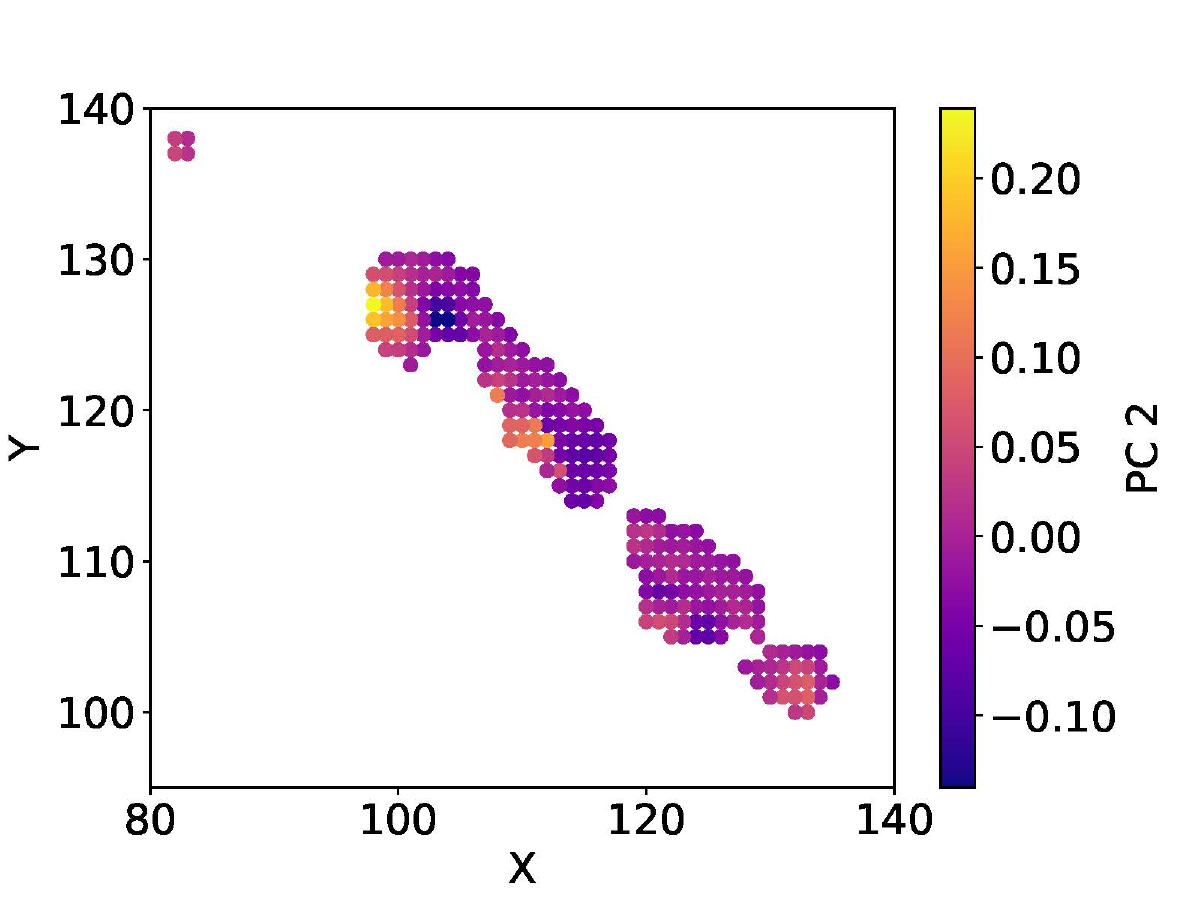}
    \includegraphics[width=0.45\textwidth]{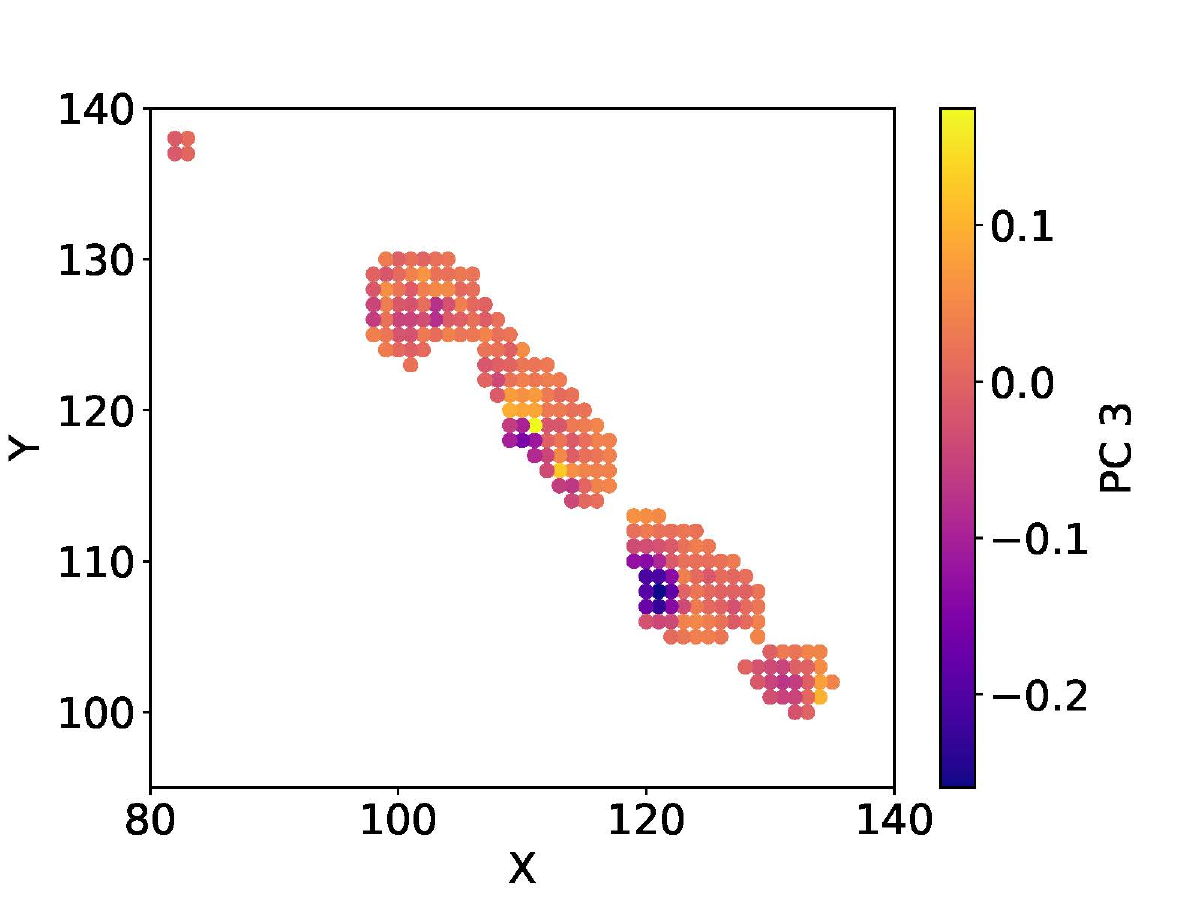}
    \caption{The 2-dimensional structure to map PC1, PC2, and PC3 obtained by the NRPCA. 
    }\label{fig:NGC253_map_PC_Doppler_corrected}
\end{figure*}

We reconstructed the 2-dimensional structure to map PC1, 2, and 3 in Fig.~\ref{fig:NGC253_map_PC_Doppler_corrected}. 
Again, PC1 represents the intensity of lines. 
In contrast to Fig.~\ref{fig:NGC253_map_PC}, new PC2 shows a more complicated, localized pattern in NGC~253. 
The map of PC3 also shows smaller scale structures. 
Recalling that PC2 suggest the existence of small-scale outflow. 
According to A17\nocite{2017ApJ...849...81A}, the spatial scale (diameter) of the most prominent region with large PC2 is $\sim 10\;\mbox{pc}$, namely it corresponds to a small-scale phenomenon, indeed. 

\begin{figure}[tb]
    \centering
    \includegraphics[width=0.36\textwidth]{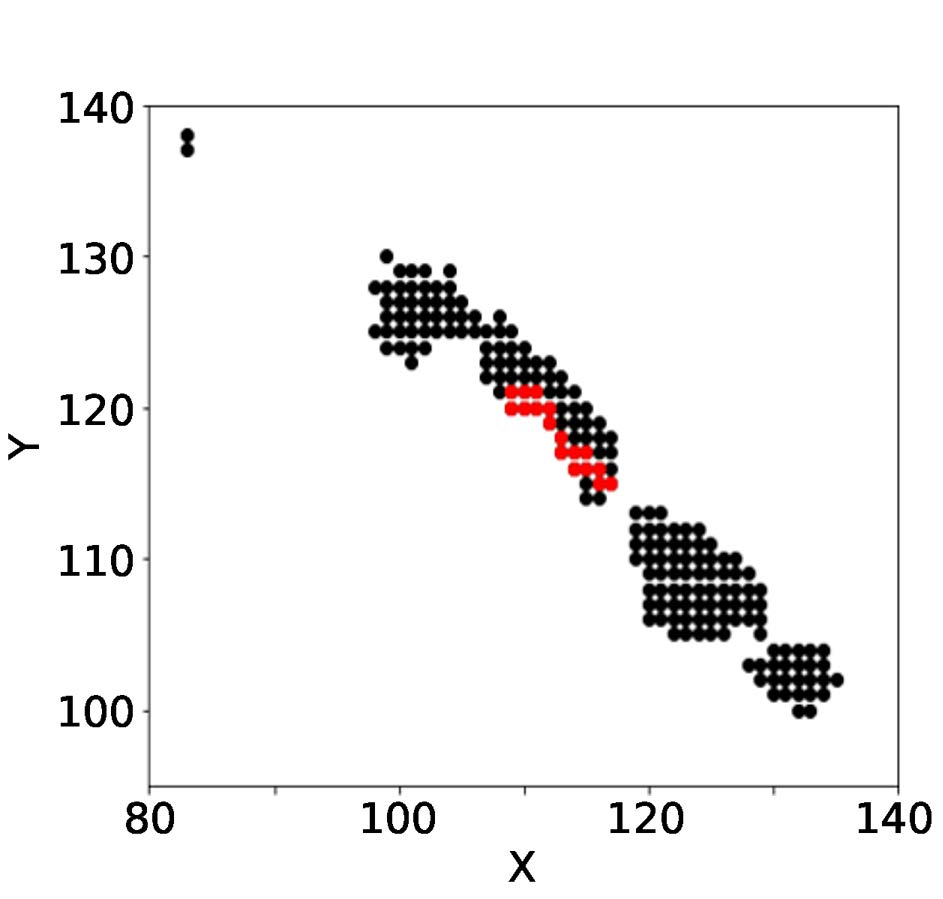}
    \caption{The region with velocity peculiarity in Fig.~\ref{fig:NGC253_scatter_PCs}.
    }
    \label{fig:NGC253_map_peculiar_velocity_region}
\end{figure}

Particularly interesting is the region with negative PC3 in Bottom panel of Fig.~\ref{fig:NGC253_map_PC_Doppler_corrected}. 
The eigenspectra show that PC3 and partially PC4 describes a local but larger-scale flow toward us, which appears in blueshift of the lines. 
\citet{2013Natur.499..450B} reported a global molecular outflow from the center of radio continuum. 
More recently, \citet{2017ApJ...835..265W} and \citet{2019ApJ...881...43K} performed detailed ALMA observations of this region. 
They made a detailed analysis and named this outflow as ``SW streamer''. 
This outflow is going out toward us, and the position of the root of the SW streamer precisely agrees with the region with negative PC3 in Fig.~\ref{fig:NGC253_map_PC_Doppler_corrected} \citep[][see their Fig.~1]{2019ApJ...881...43K}. 
The anomalous region specified by the peculiarity in velocity in Fig.~\ref{fig:NGC253_map_peculiar_velocity_region} also agrees well with the root of the SW streamer. 
In addition to the molecular observations, an outflow is also observed in H$\alpha$ and other optical emission lines by a Fabry-Perot spectroscopy \citep{2009ApJ...701.1636M}. 
Though this delineates the flow of ionized gas, the position, shape, and direction of the outflow agree fairly well with the molecular outflow. 
All these clues indicate that the high-dimensional PCA efficiently extract an outflow phenomenon from an ALMA spectroscopic map, in a purely objective way. 

\subsection{Spectral features of the Doppler-corrected data corresponding to the PCs}\label{sec:features_PCs_corrected}

\begin{figure}[t]
    \centering\includegraphics[width=0.9\textwidth]{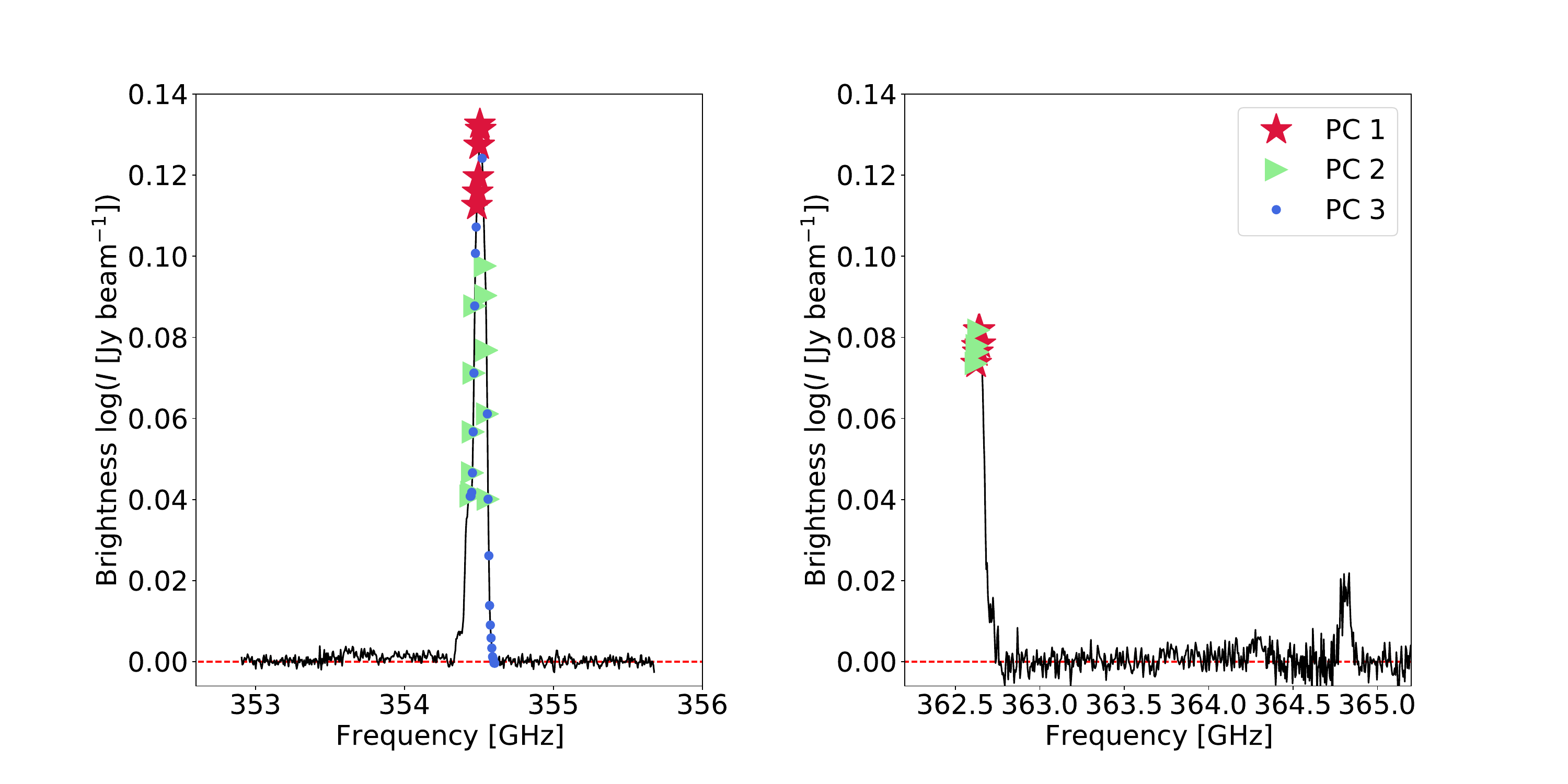}
    \caption{Responsible features to characterize PCs from the A-SPCA for the ALMA map of NGC~253, after the Doppler shift correction due to the systemic rotation.
    Stars and triangles represent PC1- and PC2-related responsible spectral features similarly to Fig.~\ref{fig:features_original}, while filled circles are PC3-related. }
    \label{fig:features_corrected}
\end{figure}

In Fig.~\ref{fig:features_corrected}, stars and triangles represent PC1- and PC2-related responsible spectral features similarly to Fig.~\ref{fig:features_original}, and filled circles are PC3-related. 
First we observe a global shift of lines compared to Fig.~\ref{fig:features_original}.
This is, of course, due to the recession velocity. 
Since the boundary between ALMA Bands are fixed at the observer's frame, some parts of the spectra go out of the boundary after the Doppler shift correction. 
As a result, we lose significant parts of the spectra in Fig.~\ref{fig:features_corrected}.
After the Doppler correction, PC1 describes the intensity of HCN(4--3) and HNC(4--3) lines more specifically, which is reflected to the fact that responsible features concentrate on the peak of these lines. 
PC2 is assigned to the part of the profile slightly far from the line center for HCN(4--3).
However, it is related to the lower-frequency side of HNC(4--3) line, though it is difficult to see clearly because this side is cut out by the correction. 
PC3 delineate further side of the higher-frequency side of HCN(4--3) line, while it has some overlap with PC2-related features on the lower-frequency side. 
These properties were already suggested from eigenspectra (Fig.~\ref{fig:eigenspectra_w_doppler_correction}), and we confirm by the detailed information from the A-SPCA. 
Thus, even if the spectra look formidably complicated, the high-dimensional PCA methods like NRPCA and A-SPCA can disentangle the information and pick up the controlling spectral features from the spectroscopic map. 
Rather surprisingly, the detailed profile of HCN and HNC lines govern almost all spectral characteristics of the ALMA map of NGC~253, including outflow phenomenon. 
Further investigation on detailed astrophysical processes from the ALMA map of NGC~253 is left to our future work.

\section{Conclusion}\label{sec:conclusion}

Evolution of galaxies is mainly driven by the star formation, a transition from ISM to stars. 
Various phases of the ISM are related, and the evolution of the ISM is a key to complete the understanding of the galaxy evolution. 
Spectroscopic observations are of vital importance to extract and interpret the information of matter in galaxies. 

Though spectroscopic mapping and integral field spectroscopy are fundamentally important to reveal the ISM physics, detailed observations are often very time-consuming, and it is not easy to map objects to obtain many independently sampled measurements. 
Consequently, we often meet a situation where the dimension of data $d$ is much larger than the sample size $n$, i.e., $n \ll d$.
This is referred to as high-dimensional low sample size (HDLSS). 
The HDLSS data are different from typical big data in astronomy (usually $n$ is huge), but $d$ is extremely large. 
In traditional studies in astronomy, the HDLSS problem has been regraded as being ill-posed and simply given up to analyze. 
However, in the end of the last millennium, a strong need of an analysis method for such data has been arisen \citep[e.g.,][]{Golub531}.
Then, during the subsequent decade, the new statistical method to tackle HDLSS problems has been vastly developed, known as the high-dimensional statistical analysis \citep[e.g.][]{aoshima2017,Aoshima2018}. 
This paper is organized twofold. 
First we introduced the general framework of high-dimensional satistical methods, and then we applied the methodology to the ALMA spectroscopic map data. 

For HDLSS data, the full information of the sample covariance is expressed by a huge $d \times d$ covariance matrix, which is implausible to handle in various aspect.  
This specific difficulty of HDLSS data is often referred to as ``the curse of dimensionality''. 
The high-dimensional statistical analysis overcomes this problem by introducing the dual representation of sample covariance matrix which is a much smaller, $n \times n$ dimension square matrix. 
The remarkable property of the dual matrix is that it contains practically the same information on the first $n$ eigenvalues of the original sample covariance matrix. 
Nevertheless, the dual sample covariance matrix is small enough to handle without the curse of dimensionality.
In addition to this most difficult issue, many unusual behaviors are known for HDLSS data are known, such as strong inconsistency, spherical concentration for Gaussian variables, and axis concentration for non-Gaussian variables. 
The high-dimensional statistics is the framework to make use of these specific aspect of the HDLSS data and construct a high-dimensional counterpart of traditional methods. 

After introducing the concept of the HDLSS data and corresponding methodology, we applied the high-dimensional version of the principal component analysis (PCA) on the NGC253 ALMA spectral map \citep{2017ApJ...849...81A}. 
ALMA mapping data are typically HDLSS in general, and as for the data in this work, $n = 231$ and $d = 1971$, clearly $n \ll d$ even though it is less extreme than the case in genomics, for example. 
The ALMA spectra are very rich in various molecular lines, and there is a very large variety in the molecular line intensities \citep{2017ApJ...849...81A}. 

We first applied the noise-reduction PCA (NRPCA) and automatic sparse PCA (A-SPCA), one of the latest methodology in high-dimensional statistical analysis, to the original ALMA map of NGC~253. 
We found that the high-dimensional PCA worked very well, and it could describe the global trend of the spectra only by the first two PCs (the contribution of the 1st and 2nd PCs is $> 50$~\%). 
PC1 approximately represents the total intensity of the spectra, and PC2 describes the Doppler shifts due to the systemic rotation of the observed region.  
Each PC consists of $\sim 5\mbox{--}20$ elements, much fewer than $d$. 
We note that practically the number of the controlling factors reduces further, since these contributing elements are part of the same spectral features.  
The controlling features were HCN(4--3) and HNC(4--3) rotational lines. 
This result strongly guarantees that the high-dimensional statistical analysis methodology works appropriately to the spectral map as HDLSS data. 

Then we applied the high-dimensional PCA to the Doppler-corrected ALMA map. 
The resulting eigenspectra of the NRPCA correspond to more specific spectral features, especially to the core or wings of the emission lines. 
While PC1 again represent the total intensity of the spectral map, PC2 appear to delineate a 10-pc scale outflow phenomenon of molecular gas. 
More impressive is that PC3 and 4 represent a larger-scale coherent flow causing a blueshift of emission lines, and PC5 the redshift. 
The region of the ALMA map where PC3 is negative concentrates around the position of the radio continuum of NGC~253. 
This map then implies that the blueshift reflects the global larger-scale outflow from the radio center of the galaxy moving toward us. 
{}To examine this, we compared the region with the observed molecular outflow reported by \citet{2019ApJ...881...43K}. 
We find that the negative PC3 region agrees well with the root of the molecular outflow, named ``SW streamer''.
This was also confirmed by the anomalous region of the velocity field in the Doppler-uncorrected map. 
Further this region also give a close agreement with the outflow of ionized gas \citep{2009ApJ...701.1636M}.
All these suggest that the PCA indeed extracted the dynamical features from the ALMA spectroscopic map, a typical HDLSS data. 

We stress that this result is not obtained by choosing a handful of features by hand, but by making use of the full information of the high-dimensional data. 
Spectroscopic maps are typical HDLSS data, and we can apply the high-dimensional statistical analysis on various unsolved problems in astrophysics.
As mentioned in \S~\ref{sec:intro}, the condition also applies to a spectroscopic mapping with satellite instruments, mainly due to the strong limitation of observation time and mission lifetime. 
The high-dimensional methodology will shed light to the spectral map of future space telescopes. 
For example, such surveys of nearby galaxies with {\sl SPICA} were discussed in details.
Though {\sl SPICA} has been cancelled, this methodology will remain useful for a future and/or possible replacement project. 
The high-dimensional statistical analysis can also be applied any type of data with small sample size and large dimension, e.g., spectroscopy of rare objects. 

\acknowledgments

We deeply thank the anonymous referee for her/his careful reading of the manuscript and constructive comments that improved the clarity of the paper much. 
This paper makes use of the following ALMA data: ADS/JAO.ALMA${\#}$2013.1.00099.S, ADS/JAO.ALMA$\#$2013.1.00735.S. 
ALMA is a partnership of ESO (representing its member states), NSF (USA) and NINS (Japan), together with NRC (Canada), MOST and ASIAA (Taiwan), and KASI (Republic of Korea), in cooperation with the Republic of Chile. The Joint ALMA Observatory is operated by ESO, AUI/NRAO and NAOJ.
This work has been supported by the Japan Society for the Promotion of Science (JSPS) Grants-in-Aid for Scientific Research (19H05076, 21H01128, 19K03937, and JP17H06130). 
This work has also been supported in part by the Sumitomo Foundation Fiscal 2018 Grant for Basic Science Research Projects (180923), the Collaboration Funding of the Institute of Statistical Mathematics ``New Perspective of the Cosmology Pioneered by the Fusion of Data Science and Physics'', and the NAOJ ALMA Scientific Research Grant Number 2017-06B. 

SC has been financially supported by the JSPS as a research fellow (DC1).
We deeply thank the {\sl SPICA} Science Team, especially members of the {\sl SPICA} the Galaxy and Nearby Galaxies Science Working Group (Fumi Egusa, Hanae Inami, Hiroyuki Kaneko, Itsuki Sakon, Yoichi Tamura, Junichi Baba, Kentaro Motohara, Yoshimasa Watanabe), to which TTT and KN belong. 
We are also grateful to Yugo Nakayama, Shiro Ikeda, and Nanase Harada for fruitful discussions.

\vspace{5mm}
\facilities{ALMA}


\software{astropy \citep{astropy2013,2018AJ....156..123A},
{\sc casa} \citep{2007ASPC..376..127M}, 
{\sc topcat} \citep{2005ASPC..347...29T}
          }



\appendix

\section{
The merit of the dual sample covariance
}
\label{sec:dual_covariance}

{
We note that the sample covariance matrix $\tilde{S}$ and the dual sample covariance $\tilde{S}_{\rm D}$ share their non-zero eigenvalues. 
We also note that the eigenvectors of $\tilde{S}$ can be calculated by the  eigenvectors of $\tilde{S}_{\rm D}$. 
See eq.~(\ref{eq:vector}). 
Thus the merit of 
the dual sample covariance ($n \times n$) is rather obvious: 
\begin{enumerate}
    \item We can reduce the computational cost significantly.
    \item We can visualize the behavior of data easier than the original data matrix ($d \times d$). 
    \item Asymptotic theory can be developed on the dual space. 
\end{enumerate}
}

\section{Geometric representations of $ \tilde{S}_{\rm D}$}
\label{sec:geometric_representation} 
{
In this section, we give geometric representations of $ \tilde{S}_{\rm D}$ when $d \longrightarrow \infty$ while $n$ is fixed \citep{yata2010,YATA2012193}. 
The geometric representations are keys to deal with the HDLSS data by “high-dimensional statistical analysis. 
}
%we basically consider the situation that $d \longrightarrow \infty$ while $n$ is fixed \citep{yata2010,YATA2012193}. 
We consider the following sphericity condition
\begin{eqnarray}\label{eq:sphericity_condition}
  \frac{\mbox{tr} \left( \tilde{\Sigma}^2\right)}{\left(\mbox{tr} \, \tilde{\Sigma}\right)^2} = \frac{\displaystyle \sum _{i=1}^d \lambda_i^2  }{\displaystyle \left( \sum _{i=1}^d \lambda_i\right)^2  } \longrightarrow 0 \quad \mbox{as} \quad d \longrightarrow \infty \; . 
\end{eqnarray}

When $\tilde{X}$ is Gaussian, \citet{10.2307/20441411}, \citet{jung2009} and YA12\nocite{YATA2012193} showed a geometric representation as
\begin{eqnarray}\label{eq:ahn_jung}
  n \left(\sum _{i=1}^d \lambda_i \right)^{-1} \tilde{S}_{\rm D} \overset{\rm P}{\longrightarrow} \tilde{I}_n  \quad \mbox{as} \quad d \longrightarrow \infty \; . 
\end{eqnarray}
{Here the $\lambda_i$s are eigenvalues of the covariance matrix $\tilde{\Sigma}$, i.e., not the sample covariance $\tilde{S}$.}
Let 
\begin{eqnarray}
  \vec{w}_i = n \left(\sum _{i=1}^d \lambda_i \right)^{-1} \tilde{S}_{\rm D} \vec{u}_i
  = n \left(\sum _{i=1}^d \lambda_i \right)^{-1} \hat{\lambda}_i \vec{u}_i 
\end{eqnarray}
and $\mathcal{S}^{n-1} \equiv \left\{ \vec{x} \in \mathbb{R}^{n} \, \middle| \; \| \vec{x}\| = 1  \right\}$: a unit $(n-1)$-sphere\footnote{YA12\nocite{YATA2012193} denoted $\mathcal{S}^{n-1}$ as $\mathbf{R}^n$. 
We used a more familiar notation in geometry in this work. }. 
YA12\nocite{YATA2012193} showed that 
\begin{eqnarray}\label{eq:surface_concentration} 
  \vec{w}_i \in \mathcal{S}^{n-1} \quad \mbox{in probability as} \quad d\longrightarrow \infty \; , 
\end{eqnarray}
that is, the surface concentration of the unit $n$-sphere. 
{}From the geometric representation by eq.~(\ref{eq:surface_concentration}), we observe that {\it the sample eigenvalues become deterministic} and converge to the same value.
This means that  it becomes impossible to estimate the eigenvalues and the eigenvectors by using $\tilde{S}_{\rm D}$ (or $\tilde{S}$) as in the conventional method. 
\begin{figure*}[tb]
    \centering
    \includegraphics[width=0.48\textwidth]{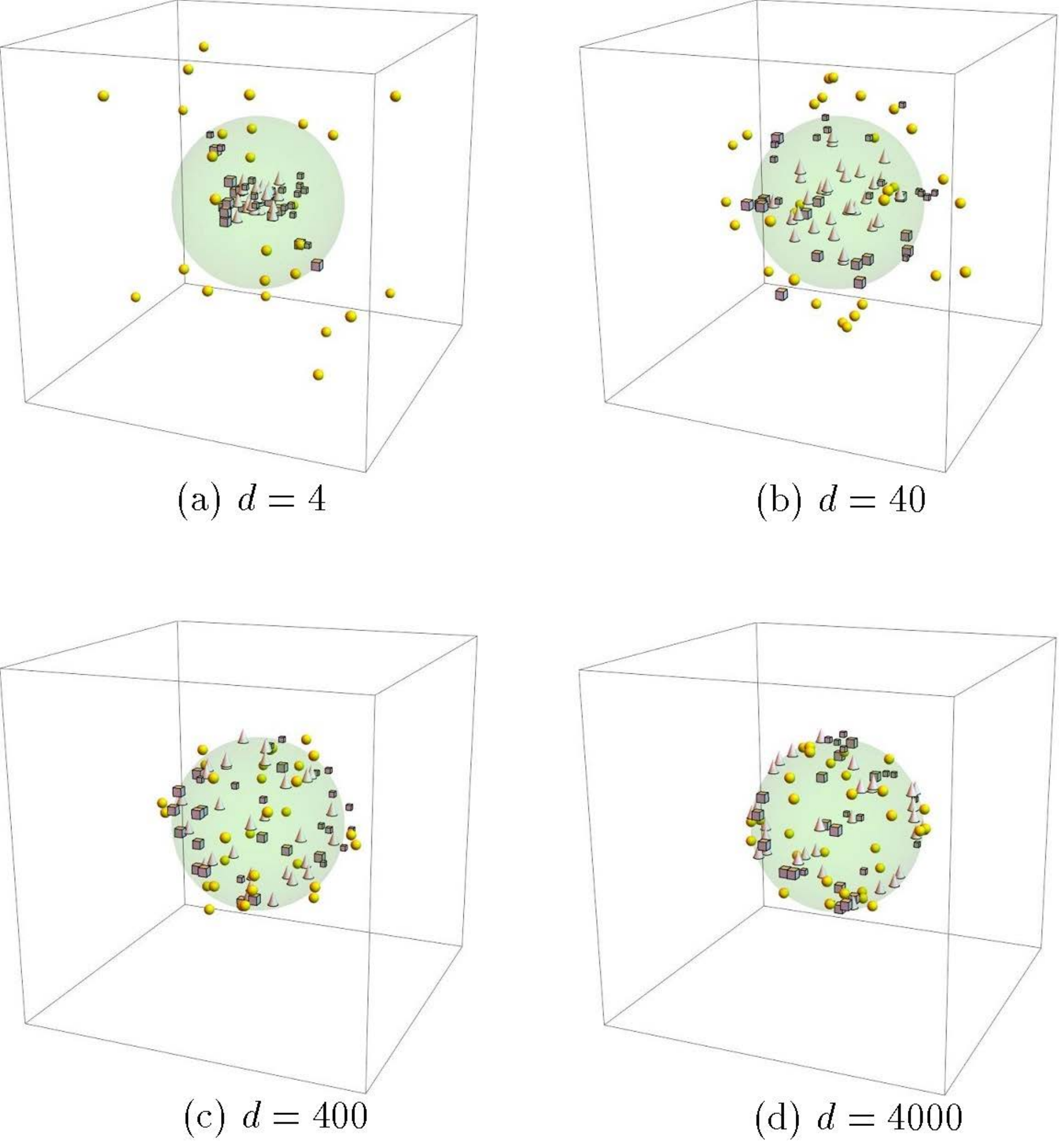}
    \quad
    \includegraphics[width=0.48\textwidth]{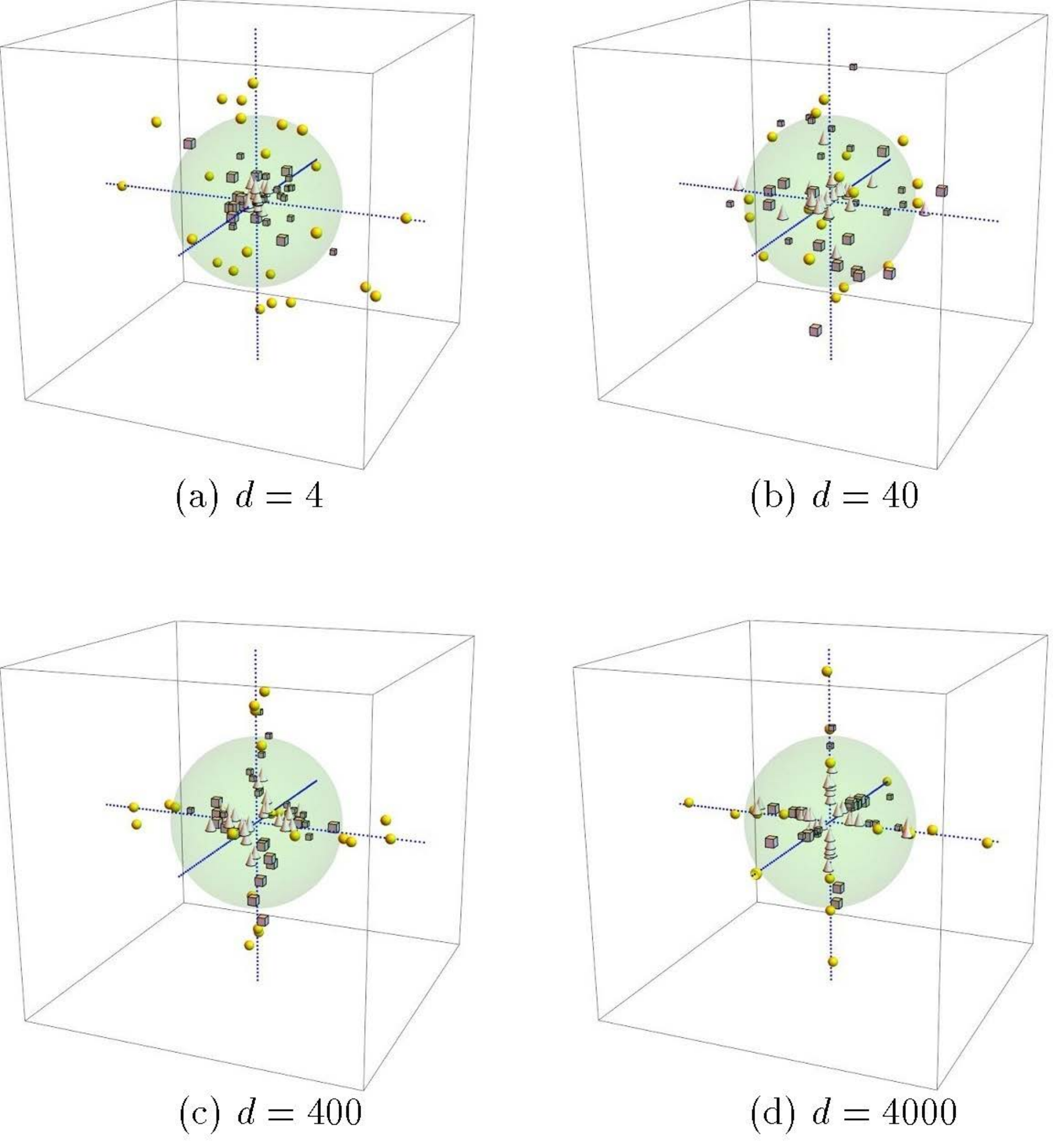}
    \caption{A sample distribution drawn from a two typical probability density distributions. Left: a sample generated from a $d$-dimensional Gaussian distribution with a mean $\vec{\mu}=\vec{0}$ and a covariance matrix $\tilde{\Sigma}_d= \tilde{I}_d$ ($\tilde{I}_d$: a $d$-dimensional identity matrix). 
    Right: a sample generated from a $d$-dimensional $t$-distribution with  $t_d(0, I_d, \nu)$, with a mean $\vec{\mu} = \vec{0}$, a covariance matrix $I_d$ and a degrees of freedom $\nu = 5$. 
    }\label{fig:high_dimension_concentration_sphere}
\end{figure*}

When $\tilde{X}$ is non-Gaussian, % and $\tilde{Z}$ is non-$\rho$-mixing,
\citet{yata2010} and YA12\nocite{YATA2012193} discovered a different type of geometric representation, 
\begin{eqnarray}\label{eq:nongaussian_representation}
  n \left(\sum _{i=1}^d \lambda_i \right)^{-1} \tilde{S}_{\rm D} - \tilde{D}_n \overset{\rm P}{\longrightarrow} \tilde{O}_n  \quad \mbox{as} \quad d \longrightarrow \infty \; . 
\end{eqnarray}
where $\tilde{D}_n$ is a diagonal matrix whose diagonal elements are of $O_{\rm P}(1)$, and $\tilde{O}_n$ is a zero matrix of dimension $n \times n$. 
Let 
\begin{eqnarray}
  \mathcal{A}^n \equiv  
  \left\{
  \left(
  \begin{array}{c}
    1 \\
    0 \\
    \vdots \\
    0
  \end{array}
  \right)
  \cup
  \left(
  \begin{array}{c}
    0 \\
    1 \\
    \vdots \\
    0
    \end{array}
    \right)
    \cup
    \dots
    \cup
  \left(
  \begin{array}{c}
    0 \\
    0 \\
    \vdots \\
    1
    \end{array}
  \right) \right\},  
\end{eqnarray}
($\mathcal{A}^n \subset{ \mathbb{R}^n}$).  
When $\tilde{X}$ is non-Gaussian, % and $\tilde{Z}$ is non-$\rho$-mixing, 
YA12\nocite{YATA2012193} showed that 
\begin{eqnarray}\label{eq:axis_concentration}
  \vec{u}_i \in \mathcal{A}^n \quad \mbox{in probability as} \quad d\longrightarrow \infty \; , 
\end{eqnarray}
that is, the axis concentration of the $n$-dimensional dual space. 
{}From the geometric representation by eq.~(\ref{eq:axis_concentration}), we can see that the eigenvectors of the data matrix converge to the unit vector of one of the axes of $\mathbb{R}^n$ and the eigenvalues would be indefinite.  

We {consider $(z_{1k}^2 -1, \dots, z_{dk}^2 -1) \; (k = 1, \dots, n)$} and its covariance matrix $\tilde{\Phi} = (\phi_{ij})$. 
YA12\nocite{YATA2012193} considered a boundary that separates the two cases described by  eq.~(\ref{eq:surface_concentration}) and eq.~(\ref{eq:axis_concentration}) as follows.
\begin{eqnarray}\label{eq:boudary}
  \frac{\displaystyle \sum _{i,j=1}^d \lambda_i \lambda_j \phi_{ij}}{\displaystyle \left( \sum _{i=1}^d \lambda_i\right)^2  } \longrightarrow 0 \quad \mbox{as} \quad d \longrightarrow \infty \; . 
\end{eqnarray}
{Note that eq.~(\ref{eq:boudary}) is equivalent to 
eq.~(\ref{eq:sphericity_condition}) under Condition~(\ref{eq:independence})}.   
Based on this, YA12\nocite{YATA2012193} proposed the following important theorem.

{\theorem {\bf [\citet{YATA2012193} I]}\label{th:YA2012}\\
Assume the sphericity condition [eq.~(\ref{eq:sphericity_condition})]. 
If the elements of $\tilde{Z}$ satisfy eq.~(\ref{eq:boudary}), we have eqs.~(\ref{eq:ahn_jung}) and (\ref{eq:surface_concentration}) as ${d} \longrightarrow \infty$. 
Otherwise, eq.~(\ref{eq:nongaussian_representation}) holds as $d \longrightarrow \infty$, and under some regularity conditions, we have eq.~(\ref{eq:axis_concentration}) as $d \longrightarrow \infty$.
}
\\

Now we are ready to examine the geometric behavior of the noise in the dual space.
Consider a sample with a $d$-dimensional Gaussian distribution with a mean $\vec{\mu} = \vec{0}$ and covariance matrix $\tilde{\Sigma}_d= \tilde{I}_d$. 
If we generate a sample $n = 3$ from this distribution, we obtain a sample distribution displayed in Fig.~\ref{fig:high_dimension_concentration_sphere}.
For a low-$d$, we have a sample distribution not very far from our intuition, i.e., the sample data concentrate on the origin and have a certain dispersion around it. 
However, for a higher-$d$, the sample data concentrate on a sphere with a radius of $(d/n)^\frac{1}{2}$ (Left panel of Fig.~\ref{fig:high_dimension_concentration_sphere}). 
This clearly shows the situation of high-dimensional data, namely the intrinsic feature of the data is completely overwhelmed by a tremendous disturbance from the huge noise sphere. 

The behavior of the sample is even more counter-intuitive for a non-Gaussian case. 
If we generate a sample from a $d$-dimensional $t$-distribution $t_d(0, I_d, \nu)$, with a mean $\vec{\mu} = \vec{0}$, a covariance matrix $I_d$ and a degrees of freedom $\nu = 5$, we observe a very particular behavior (Right panel of Fig.~\ref{fig:high_dimension_concentration_sphere}).
All the data concentrate on one of the axes for high-$d$ cases. 
This is referred to as the axis concentration, typically seen for a non-Gaussian density distribution in high dimensions. 

Thus, we understood the mathematical reason why we have the counter-intuitive behaviors of the HDLSS data in Fig.~\ref{fig:high_dimension_concentration_sphere}.
This also gives the important basis of the central method we apply in this work.

\section{Consistency for a non Gaussian case}\label{sec:non_gaussian}

{
In this section, 
we give  the consistency eq.~(\ref{eq:eigenvalues}) 
when $Z$ does not satisfy Condition.}  (\ref{eq:independence}). 
%%%
{\theorem{\bf [\citet{doi:10.1080/03610910902936083} II]}\label{th:YAII}\\
If $Z$ does not satisfy Condition (\ref{eq:independence}), eq.~(\ref{eq:eigenvalues}) holds for $i \leq m$ under the conditions
\begin{enumerate}
    \item $d \longrightarrow \infty$ and $n \longrightarrow {\infty}$ for $i$ that satisfies $\alpha_i > 1$ . 
    \item $d \longrightarrow \infty$ and $d^{2 - 2\alpha_i}/n(d) \longrightarrow 0$ for $i$ that satisfies $\alpha_i \in (0, 1]$.
\end{enumerate}
}

\noindent
Theorems~\ref{th:YAI} and \ref{th:YAII} give the condition that the estimation by the traditional PCA would have the consistency. 
{
In (2), $n$ is heavily depending on $d$ when the components of $\tilde{Z}$ do not satisfy Condition~(\ref{eq:independence}). 
For the non-Gaussian case, \citet{120002381162} proposed 
a new robust PCA called the cross-data-matrix methodology. 
}

{
\section{Comparison between traditional and high-dimensional PCA}\label{sec:comparison_PCA}
}
{
As we mentioned in the main text, high-dimensional data as the ALMA data are regarded as a sum of informative signals and noise.
Particularly for the case of HDLSS data, the traditional statistical estimation is seriously hampered by nuisance from the noise. 
In order to demonstrate the contribution of the noise sphere to the statistical estimation of an HDLSS data, we present ratios of PCA eigenvalues estimated with the traditional and high-dimensional PCA. 
For the latter, we show the estimates obtained by the NRPCA. 
As we mentioned in the main text, the eigenvalues are significantly overestimated by the traditional PCA when the data are HDLSS-type [eq.~(\ref{eq:eigenvalues_estimation})]. 
This effect can be visualized by taking a ratio of eigenvalues obtained by the traditional PCA ($\hat{\lambda}_i$) and that obtained by the NRPCA ($\check{\lambda}_i$). 
}

\begin{figure}[t]
    \centering\includegraphics[width=0.47\textwidth]{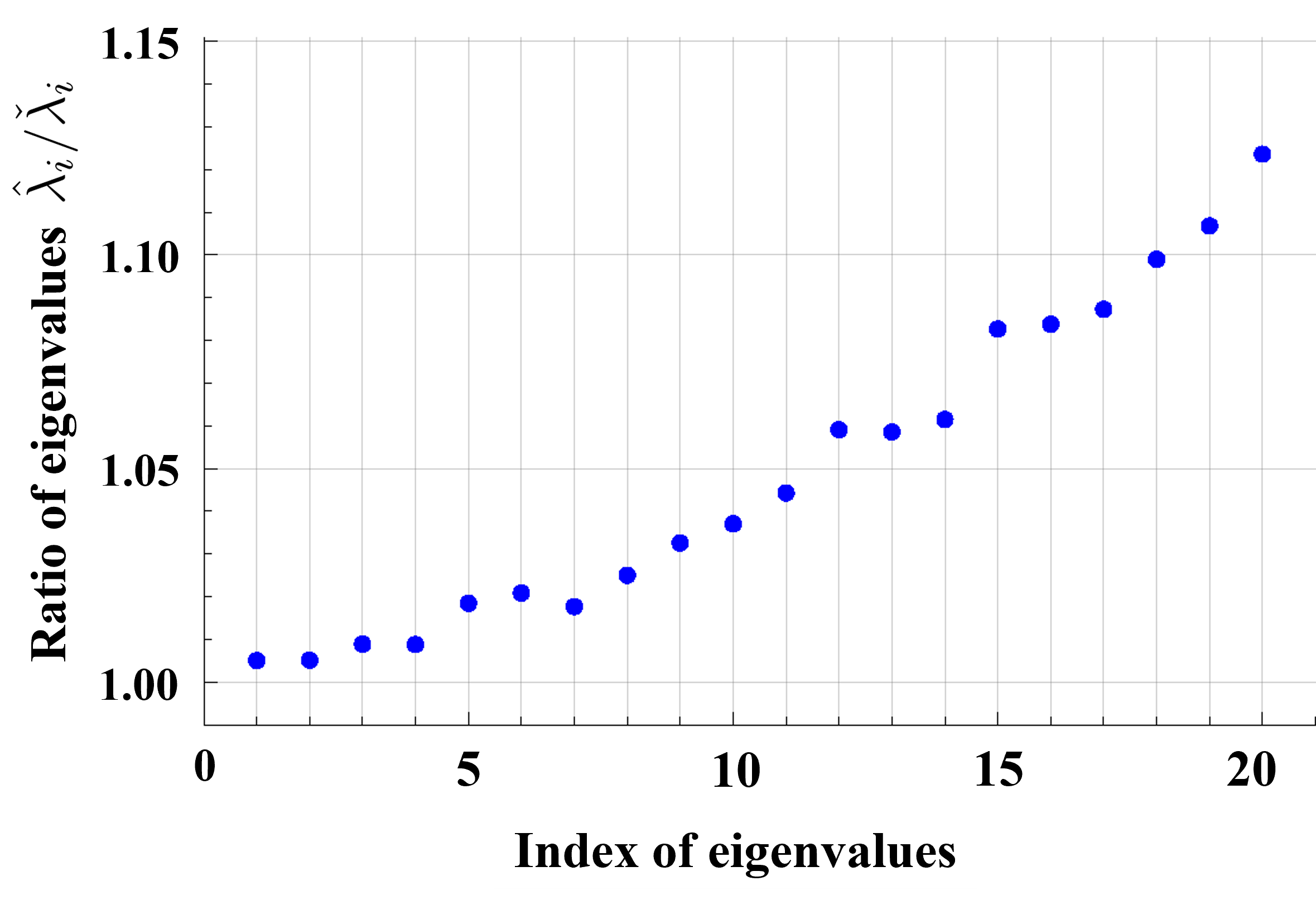}
    \centering\includegraphics[width=0.47\textwidth]{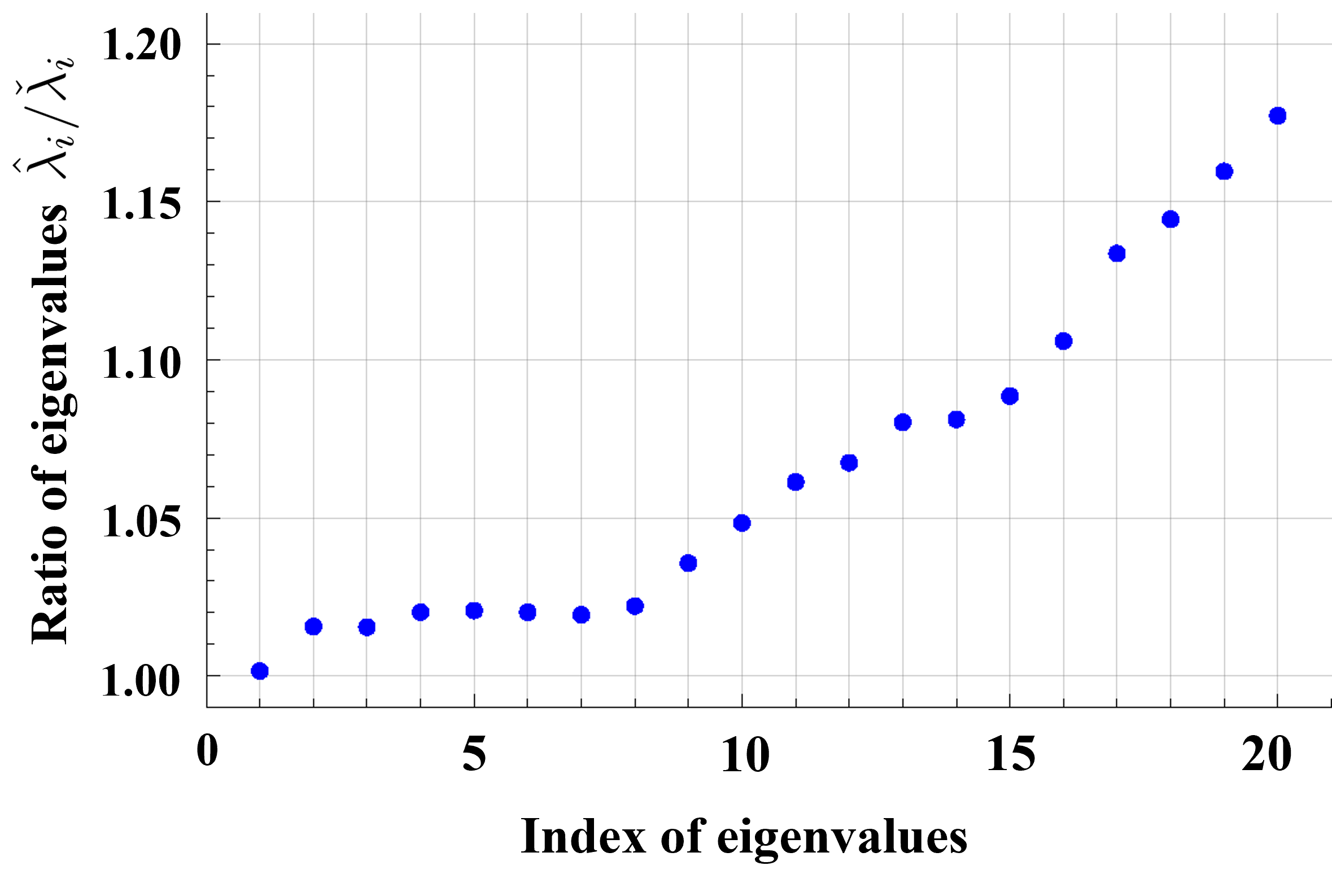}
    \caption{{Ratio of eigenvalues obtained by the traditional PCA ($\hat{\lambda}_i$) and that obtained by the high-dimensional (NR)PCA ($\check{\lambda}_i$). 
    Horizontal axes represent the index of eigenvalues, and vertical axes show the ratio $\hat{\lambda}_i/\check{\lambda}_i$ ($i = 1, 2, \dots$). 
    Left panel of Fig.~\ref{fig:comparison_traditional_highdimensional_PCA} is the result of the raw data (without Doppler velocity subtraction), while Right panel shows that of the Doppler-corrected data. 
    }}
    \label{fig:comparison_traditional_highdimensional_PCA}
\end{figure}

{
The result is presented in Fig.~\ref{fig:comparison_traditional_highdimensional_PCA}. 
Left panel of Fig.~\ref{fig:comparison_traditional_highdimensional_PCA} is the result of the raw data (without Doppler velocity subtraction), while Right panel shows that of the Doppler-corrected data. 
In both cases, it is clearly shown that traditionally estimated eigenvalues are systematically larger than NRPCA ones. 
We should note that the contribution of noise almost monotonically increases for higher-order eigenvalues, both for the raw and Doppler-corrected datasets. 
This is because the eigenvalues are smaller for higher-order indices with respect to the noise contribution [cf.\ eq.~(\ref{eq:eigenvalues_estimation}].
}

{
We also note that, since the data dimension $d$ is not extremely higher than data size $n$, the overestimation of the eigenvalues are relatively small. 
We stress, however, all the values are overestimated due to the additional contribution from noise. 
Namely, for forthcoming data that have a much stronger HDLSS property as that of ALCHEMI, this bias will become much larger. 
Thus, this examination shows the potential power of the high-dimensional PCA, and high-dimensional statistical analysis in general.
}

~

\section{Doppler shift correction of a line: example}\label{sec:Doppler_shift_correction}

\begin{figure*}[tbh]
    \centering
    \includegraphics[width=0.9\textwidth]{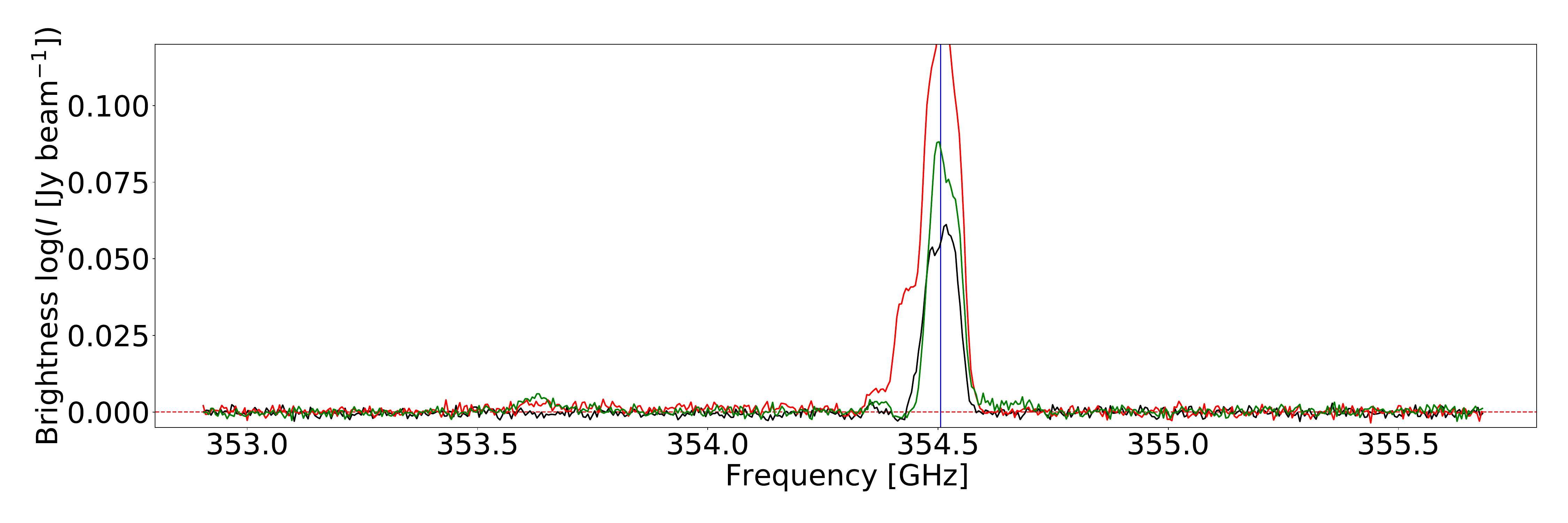}
    \caption{Example of the Doppler shift-corrected molecular emission line, HCN(4--3). 
    Red, green, and blue are the corrected spectra so that the line center agree with its restframe frequency.
    }\label{fig:NGC253_shifted_emission}
\end{figure*}

We show an example of the Doppler shift correction, HCN(4--3) line in Fig.~\ref{fig:NGC253_shifted_emission}
Red, green, and blue are the corrected spectra from different spatial position on the map of NGC~253, and we see that the line center agree with its restframe frequency.
Since each line has a complicated profile, the line center was determined by the Gaussian fitting. 
The blue vertical solid line represents the restframe frequency of the HCH(4--3) line, 354.5~GHz.

\bibliography{high_dimension_NGC253_rev3}{}
\bibliographystyle{aasjournal}

\end{document}